\documentclass[fleqn,usenatbib]{mnras}

\usepackage{newtxtext,newtxmath}

\usepackage[T1]{fontenc}

\DeclareRobustCommand{\VAN}[3]{#2}
\let\VANthebibliography\thebibliography
\def\thebibliography{\DeclareRobustCommand{\VAN}[3]{##3}\VANthebibliography}

\usepackage{graphicx}	
\usepackage{amsmath}	
\usepackage{orcidlink}
\usepackage{subcaption}

\newcommand{\rps}{\ensuremath{r_{\textsc{ps}}}}
\newcommand{\ips}{\ensuremath{i_{\textsc{ps}}}}

\newcommand{\wps}{\ensuremath{w_{\textsc{ps}}}}
\newcommand{\grizy}{\ensuremath{grizy_{\textsc{ps}}}}

\title[The gravitational wave candidate S251112cm]{The gravitational wave candidate S251112cm: near-complete optical coverage with ATLAS, Pan-STARRS and the Vera C. Rubin Observatory}

\author[A.~J.~Smith et al.]{A.~J.~Smith$^{1}$\orcidlink{0009-0007-9351-675X},\thanks{E-mail: asmith67@qub.ac.uk}
M.~Nicholl$^{1}$\orcidlink{0000-0002-2555-3192},
S.~J.~Smartt$^{2,1}$\orcidlink{0000-0002-8229-1731},
J.~H.~Gillanders$^{2}$\orcidlink{0000-0002-8094-6108},
D.~R.~Young$^{1}$\orcidlink{0000-0002-1229-2499},
K.~W.~Smith$^{2,1}$\orcidlink{0000-0001-9535-3199},
\newauthor
S.~P.~MacBride$^{3}$\orcidlink{0000-0002-9514-7245},
D.~Magill$^{1}$\orcidlink{0009-0000-6521-8842},
C.~R.~Angus$^{1}$\orcidlink{0000-0002-4269-7999},
T.-W.~Chen$^{4}$\orcidlink{0000-0002-1066-6098},
M.~R.~Alarcon$^{5,6,7}$,
T.~de Boer$^{8}$\orcidlink{0000-0001-5486-2747},
\newauthor
K.~C.~Chambers$^{8}$\orcidlink{0000-0001-6965-7789},
M.~E.~Huber$^{8}$\orcidlink{0000-0003-1059-9603},
C.-C.~Lin$^{8}$\orcidlink{0000-0002-7272-5129},
T.~B.~Lowe$^{8}$\orcidlink{0000-0002-9438-3617},
G.~S.~H.~Paek$^{8}$\orcidlink{0000-0002-6639-6533},
X.~Sheng$^{1}$\orcidlink{0000-0002-6527-1368}
\\
$^{1}$Astrophysics Research Centre, School of Mathematics and Physics, Queen's University Belfast, Belfast BT7 1NN, UK \\
$^{2}$Department of Physics, University of Oxford, Keble Road, Oxford, OX1 3RH, UK\\
$^{3}$Physik-Institut, University of Zurich, Winterthurerstrasse 190, 8057 Zurich, Switzerland\\
$^{4}$Graduate Institute of Astronomy, National Central University, 300 Jhongda Road, 32001 Jhongli, Taiwan\\
$^{5}$Light Bridges, Observatorio Astronómico del Teide. Carretera del Observatorio del Teide, s/n, E-38570 Güímar, Canary Islands, Spain\\
$^{6}$Instituto de Astrofísica de Canarias (IAC), C/ Vía Láctea s/n, E-38205 La Laguna, Canary Islands, Spain\\
$^{7}$Departamento de Astrofísica, Universidad de La Laguna, E-38206 La Laguna, Canary Islands, Spain\\
$^{8}$Institute for Astronomy, University of Hawai`i, 2680 Woodlawn Drive, Honolulu HI 96822, Hawai`i\\
}

\date{Accepted XXX. Received YYY; in original form ZZZ}

\pubyear{\the\year{}}

\begin{document}
\label{firstpage}
\pagerange{\pageref{firstpage}--\pageref{lastpage}}
\maketitle

\begin{abstract}
On 2025 November 12, the LIGO-Virgo-KAGRA (LVK) network detected S251112cm, a compact binary merger candidate with a false alarm rate of one per four years at a luminosity distance of $93\pm27$\,Mpc. The probability that the lighter component's mass lies below $1\,\text{M}_\odot$ is $>99$\%, while the probability that either component lies at $1\text{-}3\,\text{M}_\odot$ is only 8\%, leaving the nature of the progenitor system unclear. We present Pan-STARRS and ATLAS optical follow-up of the 90\% localisation region, and quantify Vera C. Rubin Observatory target-of-opportunity observations. In the North, Pan-STARRS imaged 1084\,deg$^2$ (65\%) within 7 days to a median depth of 21.9 mag and ATLAS 1545\,deg$^2$ (87\%) within 14 days to 19.3 mag across both hemispheres; in the South, Rubin covered 849\,deg$^2$ (36\%) 2--10 days post-merger to median depths of 24.5 ($g$) and 23.5 ($i$) mag. They reported 173, 93 and 67 candidates respectively (269 unique), which we assess against AT~2017gfo-like kilonova and SN~1998bw-like broad-lined Type Ic supernova models using prior activity, host luminosity distance, photometric evolution and absolute magnitude. Eight survive, mainly for lack of distance information. Six of these lie in hosts fainter than any in a volume-limited comparison sample, suggesting they are background sources. AT~2025aimy declines too slowly for the AT~2017gfo-like model, and a VLT host spectrum places AT~2025alqe outside the LVK volume, leaving no convincing counterpart candidates. Pan-STARRS and Rubin together exclude both models in $\sim$80\% of the 90\% localisation volume: the deepest wide-area constraint on a subsolar-mass merger to date and, to our knowledge, the most complete optical coverage of a significant subsolar-mass candidate.
\end{abstract}


\section{Introduction}
\label{sec:introduction}

Gravitational wave (GW) astronomy has rapidly become a cornerstone of time-domain astrophysics, enabling the detection of compact binary mergers involving binary neutron stars (BNS), binary black holes (BBH), and neutron star-black hole (NSBH) systems \citep{LVK_GWTC-4.0}. This progress has been driven by the advanced detector network of the LIGO-Virgo-KAGRA (LVK) Collaboration \citep{aasi-2015,accadia-2012, Acernese2015,akutsu-2020}, whose increasing sensitivity has led to a growing sample of confidently detected events \citep{LVK_GWTC-4.0}. While GW observations alone provide robust constraints on source properties such as component masses, spins, and luminosity distance, the process of finding an electromagnetic (EM) counterpart is essential for fully characterising the physical processes at play \citep{Metzger2017_kilonovae}. 

BNS mergers are expected to unbind ejecta through disc wind, tidal stripping and shocks; this gives rise to kilonova emission powered by the radioactive decay of r-process elements \citep{Metzger2010, Roberts2011, Kansen2017}. In NSBH systems, the presence of ejecta is conditional on whether the neutron star (NS) is tidally disrupted outside the black hole event horizon. This depends on the black hole mass, spin, and the NS equation of state (EOS) \citep{Foucart2012, Kawaguchi2016}. 

The detection of GW170817 marked a transformative period in astrophysics, representing the first direct multimessenger observation of a compact binary merger through both GW and EM signals \citep{Abbot2017_discovery, Abbott_2017_MMA_Summary}. The GW signal, consistent with a BNS inspiral, was accompanied by a short gamma-ray burst (sGRB) and the subsequent discovery of the optical transient AT~2017gfo at a luminosity distance of 40$^{+8}_{-14}$\,Mpc \citep{Abbot2017_discovery}. This firmly established kilonovae as EM counterparts of NS mergers. GW170817 triggered an extensive global follow-up campaign. Wide-field survey observations played a key role in identifying and characterising the transient in its early phases \citep{SoaresSantos_2017, tanvir-2017, Lipunov2017_170817}, while targeted follow-up observations provided detailed spectroscopic and multi-wavelength constraints on the ejecta composition, velocity structure, and radiative processes \citep{Coulter2017_170817, valenti-2017, Arcavi_170817, chornock-2017, nicholl-2017, Smartt2017, pian-2017, tanaka-2017, Drout2017_170817, Kasliwal2017_270817, Shappee2017_170817, Troja2017_170817}. Together, these observations provided unprecedented insight into r-process nucleosynthesis, relativistic outflows, and the structure of merger ejecta \citep{Margutti_2021}.

Following GW170817, follow-up campaigns have targeted subsequent GW alerts, although no confirmed optical counterparts have been identified, including for the likely BNS merger GW190425 \citep{Abbott_2020_GW190425_Discovery}. Its missing counterpart was likely due to the event’s large localisation region, 8284\,deg$^2$ (a significant fraction of which was in the day time),  and greater distance (159$^{+69}_{-72}$\,Mpc) \citep{Abbott_2020_GW190425_Discovery}, highlighting the continued challenges of rapid EM counterpart identification and the importance of wide-field, deep, and high-cadence surveys. Several NSBH mergers have also been detected in GW, though none has yielded a confirmed EM counterpart to date. In most cases the mass ratio was large and the black hole spin consistent with zero, so the neutron star was likely swallowed whole rather than tidally disrupted, leaving little or no material to power a kilonova \citep{Gompertz_2022, 2021Abbott}. For a review of the BNS and NSBH sources followed up between 2016 and 2025 during the LVK Collaboration's observing runs O2--O4, see \citet{Nicholl2025_review}.

Wide-field facilities including ZTF \citep{Bellm2019_ZTF, Graham2019_ZTF, Anand2021, Ahumada2024_ztf_O4, Kasliwal2020}, ATLAS \citep{Tonry2018_ATLAS, Smartt2016_GW151226, Gillanders2025_250818k}, Pan-STARRS \citep{Chambers2016arXiv_PanSTARRS1, Smartt2016_GW150914, Smartt2016_GW151226, Gillanders2025_250818k, Fulton2025}, WFST \citep{Wang2023_WFST, Liu2023_WFST3, Liang2024_WFST2}, ASAS-SN \citep{deJaeger2022_ASAN_SN}, BlackGEM/MeerLICHT \citep{deWet2021, Groot2024}, DECam \citep{Flaugher2015_DECam, Zhang2025}, KNTraP \citep{VanBemmel2025_KNTraP}, and GOTO \citep{Steeghs2022_GOTO, Lyman2026_GOTO} have provided rapid, deep imaging of GW localisation regions to identify transient candidates. Other rapid-response efforts have supported
targeted follow-up, including GECKO \citep{Paek2024_190425, Paek2025_230518}, J-GEM \citep{Takahashi2026_JGEM}, SAGUARO \citep{Lundquist_SAGUARO_O3, Paterson_2021_SAGUARO, Hosseinzadeh2024_SAGUARO}, GRANDMA \citep{Agayeva2022_GRANDMA}, and Kinder at Lulin Observatory \citep{LOT_GCN14}. Additionally, spectroscopic classification and candidate vetting have been supported by collaborations such as ePESSTO+ \citep{Smartt2015_PESSTO, Smartt2016_GW150914}, the Las Cumbres Observatory network \citep{Arcavi2017_LCO_GW, Keinan2026_LCO_GW}, and GROWTH \citep{Andreoni2020_GROWTH, Coughlin2019}.

Looking ahead, the next generation of observatories is expected to significantly enhance the prospects for detecting EM counterparts to GW events, in particular the Vera C. Rubin Observatory with its 8.4\,m (6.5\,m effective) primary mirror and 9.6\,deg$^2$ field of view \citep{Ivezic2019_rubin}. Rubin will be able to achieve single-visit depths of $\sim 24.5$\,(AB) in $r_{\textsc{LSST}}$ \citep{Ivezic2019_rubin}. 

In addition to BNS, NSBH and BBH merger templates, in O4 LVK introduced low-latency searches for compact binary mergers which contain at least one subsolar-mass (SSM) object with mass in the range $\sim 0.2$--$1$ M\,$_{\odot}$ in their signal pipelines \citep{Hanna2025, Tsukada2023, Allene_2025}. Objects in this mass range are difficult to explain through standard stellar evolution: accretion-induced and neutrino-driven collapse struggle to explain remnant NS masses less than $\sim$1.2\,M$_{\odot}$ \citep{Suwa2018, Muller2025_NuetrinoDriven_SN}. Therefore, more speculative formation channels, such as fragmentation during protoneutron star formation or neutron star collisions, may need to be considered alongside the possible existence of primordial black holes \citep{Horowitz2010_SSMNS_GW, Baumgarte2026_PBH, Haque2026}.

Some of these channels, if realised in nature, could also produce EM emission. One such possibility is in a compact binary merger with at least one SSM NS. In this case, the merger could still eject neutron-rich material powering a kilonova.  Alternatively, during a collapsar event (core collapse in a rapidly rotating massive star), the environment in the resulting accretion disc could give rise to cooling-induced fragmentation forming multiple SSM NSs, which could undergo sequential mergers producing a GW signal with a SSM component \citep{Metzger2024_SSM_Ic_BL, Chen_and_Metzger_SSM_Ic_BL}. In this case, the EM emission would be expected to precede the GW signal, and be dominated by the emission from the young supernova rather than a standard kilonova \citep{Metzger2024_SSM_Ic_BL}.

The first example of a potential SSM merger with EM follow-up is S250818k, a sub-threshold signal initially assigned a 71\% probability of terrestrial origin \citep{S250818k_discovery_GCN}. Although its false alarm rate remained high, the parameters inferred under the assumption of an astrophysical origin were consistent with a neutron star-bearing merger, motivating follow-up. Notably, this included a low chirp mass which implied at least one of the binary components had a mass $<$1\,M$_\odot$. Follow-up did not yield a confirmed counterpart. However, the observations revealed and classified a Type IIb supernova within the skymap, the connection of which to the GW source has been debated \citep{Kasliwal2025_250818k, Gillanders2025_250818k, ENGRAVE_250818k, Hall2026_ZTF_S251112cm}. The absence of an unambiguous EM counterpart reiterates the importance of rapid follow-up campaigns when searching for counterparts to compact object mergers.  

Here we present optical follow-up of S251112cm, the most significant SSM merger signal reported by LVK to date. In this work we assess wide-field Pan-STARRS and ATLAS follow-up, in addition to quantifying the public Vera C. Rubin target-of-opportunity (ToO) observations. Complementary studies reporting EM follow-up and analysis of S251112cm have been published by a number of teams. \citet{Hall2026_ZTF_S251112cm} report ZTF, FTW, and DECam coverage of 56\% of the localisation region and identify SN~2025adtq (discovered by Pan-STARRS) as a Type IIb supernova within the localisation region. \citet{Paek2026} present a two-month, galaxy-targeted $J$-band campaign with the United Kingdom Infrared Telescope (UKIRT), monitoring 59 candidate hosts to typical depths of $J \sim 22$--23\,mag with no convincing transient detected. \citet{Liu2026} report WFST coverage of $\sim$780\,deg$^2$, or 51\% of the updated localisation probability, rising to $\sim$67\% when combined with DECam, and identify no viable rapidly evolving counterpart. \citet{Vieira2026_TROVE} report galaxy-targeted, photometric and spectroscopic follow-up and introduce TROVE, the Tool for Rapid Object Vetting and Examination (Franz et al., in preparation), which scores candidates in real time to assist with follow-up allocation during GW triggers. 

\begin{figure*}
    \centering
    \includegraphics[width=0.88\textwidth]{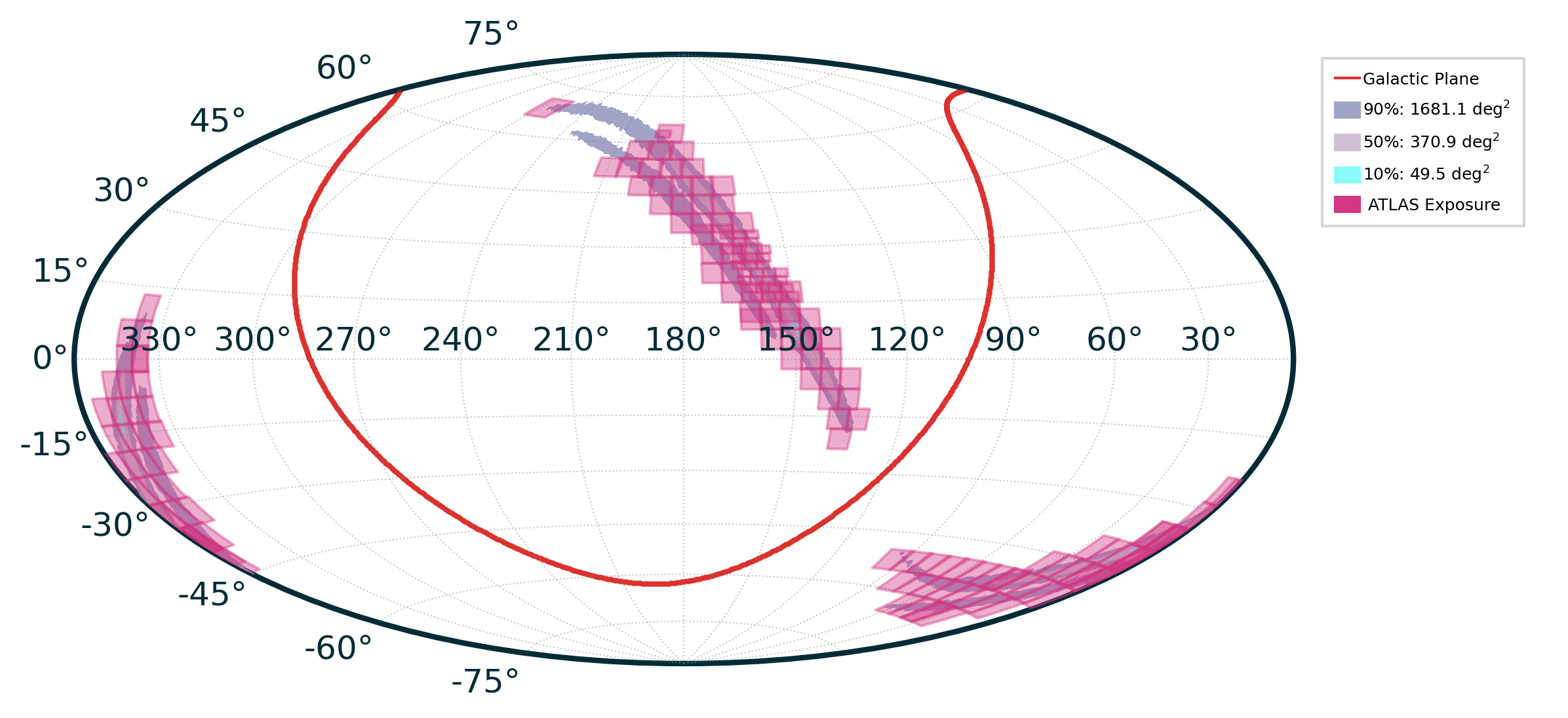}
    \includegraphics[width=0.88\textwidth]{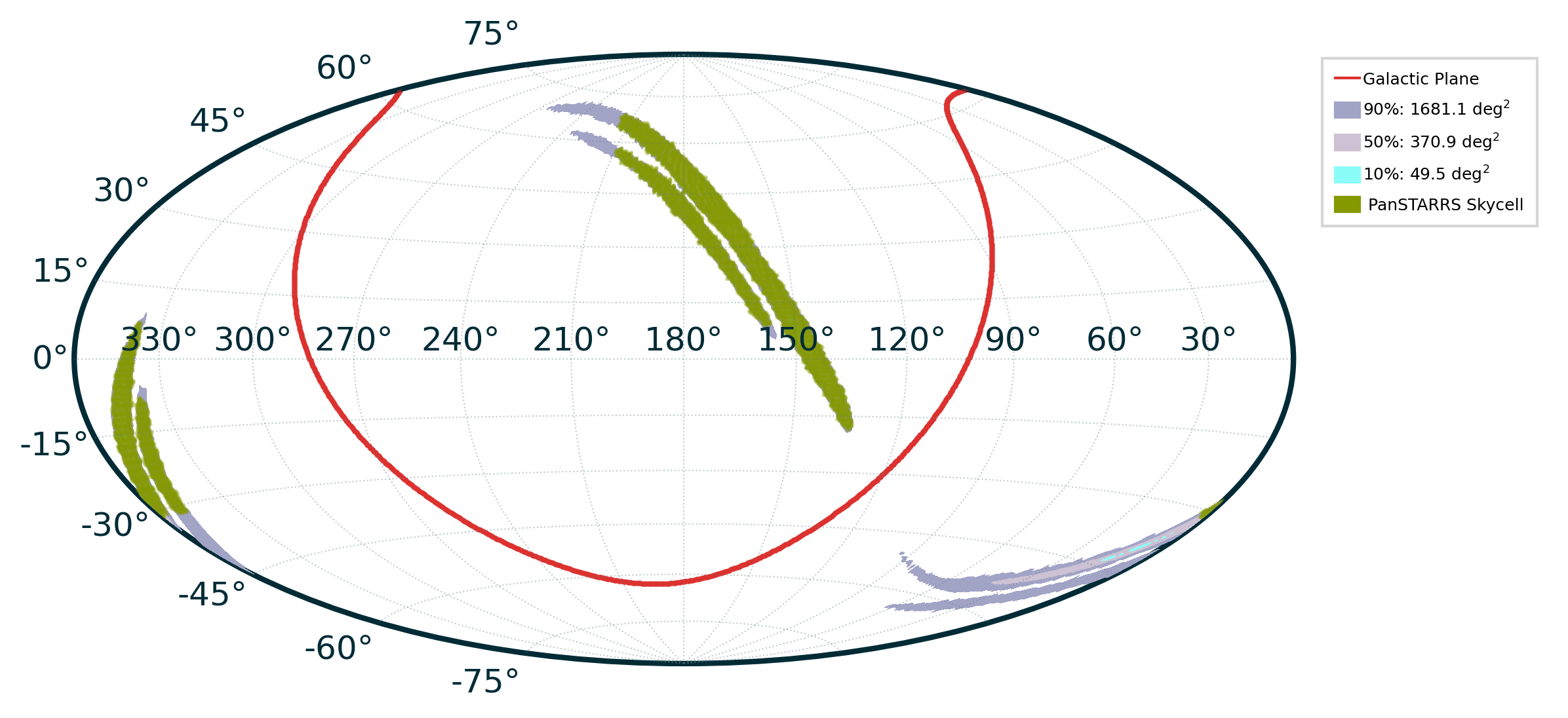}
    \includegraphics[width=0.88\textwidth]{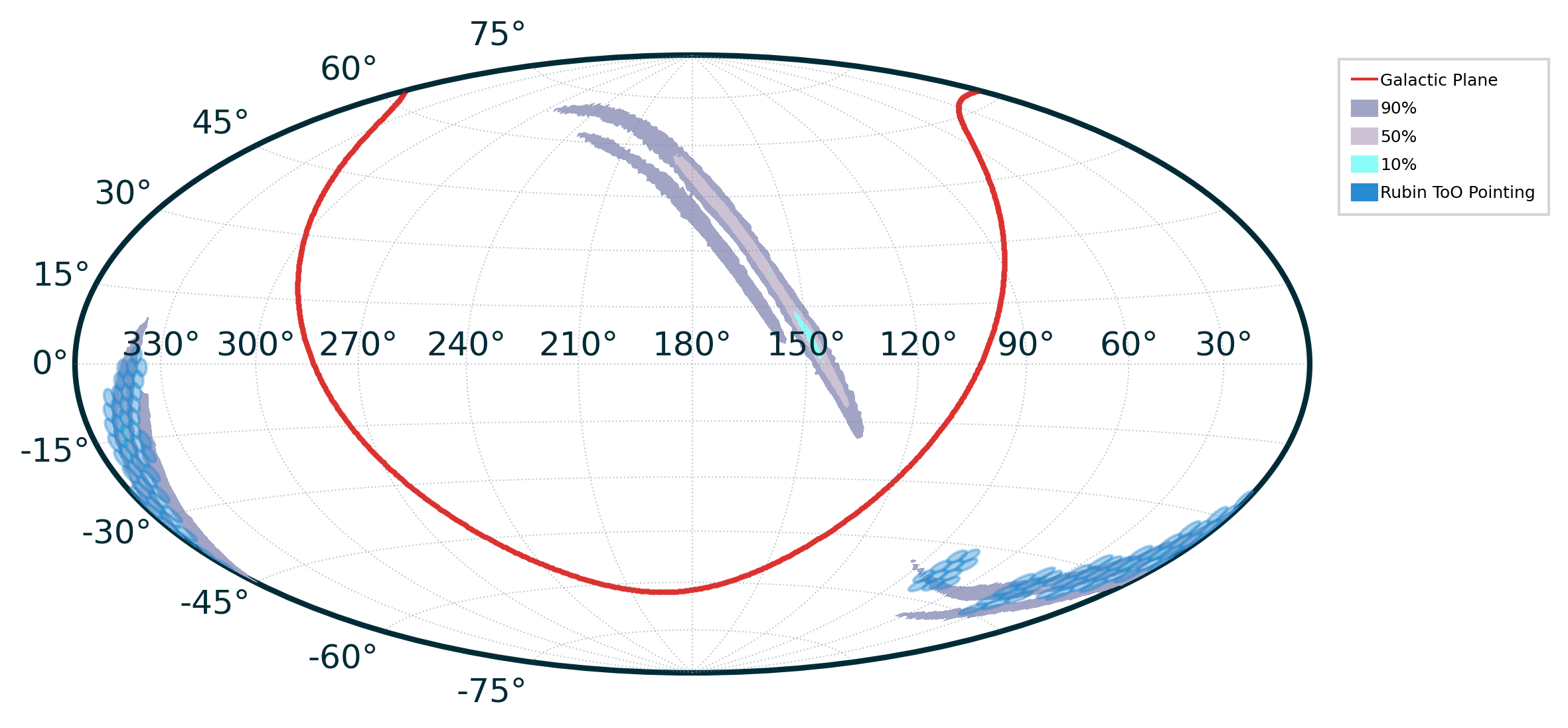}
    \caption{Coverage of the S251112cm localisation region by ATLAS (top), Pan-STARRS (middle), and the Rubin Observatory (bottom). Shaded contours mark the 90\% (1681.1\,deg$^2$), 50\% (370.9\,deg$^2$), and 10\% (49.5\,deg$^2$) credible regions of the \texttt{bilby.multiorder.fits} skymap, and the red line traces the Galactic plane. Coloured tiles show the individual ATLAS exposures (5.4$^\circ \times$ 5.4$^\circ$ for the standard units and $\sim$3.3$^\circ \times$ 2.2$^\circ$ for each Teide module), Pan-STARRS skycells ($\sim$0.4$^\circ \times$ 0.4$^\circ$ stacking tiles), and Rubin ToO pointings (single LSSTCam visits of 9.6\,deg$^2$), covering 1545, 1084, and 849\,deg$^2$ of the 90\% region, respectively.
    }
    \label{fig:skymap}
\end{figure*}

This paper is structured as follows. In Section \ref{sec:observations} we outline the GW observations of S251112cm carried out by \citet{LVGCNdiscovery} alongside the optical coverage of the skymap with the ATLAS, Pan-STARRS and Vera C. Rubin observatories. In Section \ref{sec:counterparts_and_vetting} we outline our models for an AT~2017gfo and SN~1998bw-like kilonova and Type Ic-BL supernova respectively using historical observations, and we present a framework for determining whether a candidate is associated with an SSM merger using a combination of host properties and time-series photometric data. In Section \ref{sec:results} we quantify the outcome of this vetting process and, where redshifts are available, constrain the host galaxy distances of the surviving candidates. We then assess how effectively the ATLAS, Pan-STARRS and Rubin observations exclude an AT~2017gfo or SN~1998bw-like event within their respective covered fractions of the skymap.

Throughout this work, we assume a standard $\Lambda$CDM cosmology: H$_{0}$ = 70 km s$^{-1}$\,Mpc$^{-1}$, $\Omega_{M}$ = 0.3, $\Omega_{\Lambda}$ = 0.7.

\section{Observations}
\label{sec:observations}

\subsection{S251112cm}
\label{sec_sub:S251112cm_obs}
S251112cm was detected by the LIGO Hanford, LIGO Livingston, and Virgo observatories on MJD 60991.638025 (2025 November 12 15:18:45.363 UTC) and announced publicly on MJD 60991.663484 with a false-alarm rate (FAR) of one in four years at a distance of 93 $\pm$ 27\,Mpc \citep{LVGCNdiscovery,GCN_LVK_2025b}. The standard classification probabilities for BNS, NSBH and BBH binaries vs terrestrial noise, and for the merger having matter outside of the final compact object (HasRemnant), have not been calculated for this candidate. Both are produced only by searches targeting stellar-mass binaries: because there is no established astrophysical population of subsolar-mass compact objects, the low-latency subsolar-mass searches assign significance by false alarm rate alone rather than computing a probability of astrophysical origin \citep{Allene_2025}. Provided S251112cm is of astrophysical origin, the probability that the mass of the lighter object in the binary is $<$1M\,$_{\odot}$ is $>99$\%, and the probability that at least one of the binary objects is a NS with mass between 1--3\,M$_\odot$ is 8\% \citep{LVGCNdiscovery}. Additionally, the probability that either of the compact objects has a mass within the range 3--5\,M$_\odot$ is \textless1\% and the source chirp mass lies in the range 0.1--0.87 M\,$_\odot$.  

While the distance and FAR both met the Pan-STARRS and ATLAS trigger criteria, without a BNS + NSBH probability $\ge90$\%, neither was automatically triggered \citep{Smartt2024_GW190425}. However, due to the uniqueness of the event we opted to trigger targeted Pan-STARRS follow-up of the skymap (discussed in Section \ref{sec_sub:Pan-STARRS_obs}). With its larger field-of-view and higher cadence, ATLAS would cover the majority of the skymap during routine operation (discussed in Section \ref{sec_sub:ATLAS_obs}). 

A number of other teams also conducted follow-up; results from these searches were published on the General Coordinates Network (GCN) and the Transient Name Server (TNS). Optical results from GOTO and BlackGEM/MeerLICHT identified five transients within the localisation region within $\sim$1 day post GW, two of which were shown to have detections prior to the signal by ATLAS \citep{GOTO_GCN1, BlackGEM_GCN2, ATLAS_GCN_3}. Within the following weeks, additional candidates were reported by FTW, ZTF, DECam, GRANDMA, the Vera C. Rubin Observatory, WFST, J-GEM, SVOM/VT, Kinder/LOT, GECKO/KMTNet, and ATLAS \citep{FTW_GCN4, ZTF_GCN5, ATLAS_GCN6, DECam_GCN7, GRANDMA_GCN8, GRANDMA_GCN9, Rubin_GCN10, Rubin_GCN16, WFST_GCN11, J_GEM_GCN12, SVOM_VT_GCN13, LOT_GCN14, GECKO/KMTNet_GCN15}. At higher energies, the Fermi Gamma-ray Space Telescope, Glowbug, and MAXI observed the localisation region within hours of the LVK alert but found no convincing counterparts, while follow-up from the Neil Gehrels Swift Observatory and the Einstein Probe over the first $\sim$2 days post-merger identified several X-ray candidates for further investigation \citep{Fermi_GCN17, glowbug_GCN, MAXI_GCN18, SWIFT_GCN19, EP_FXT_GCN20}.

\subsection{Pan-STARRS}
\label{sec_sub:Pan-STARRS_obs}
Pan-STARRS (PS) is a twin 1.8-m telescope system (Pan-STARRS1 and Pan-STARRS2), both situated atop Haleakala mountain on the Hawaiian island of Maui \citep{Chambers2016arXiv_PanSTARRS1}. Pan-STARRS1 (PS1) has a 1.4 gigapixel camera and the 0.26 arcsec pixels provide a focal plane with a diameter of 3.0 degrees, and a field-of-view (FOV) area of 7.06 deg$^{2}$. The Pan-STARRS2 (PS2) telescope hosts a 1.5 gigapixel camera and supports a slightly larger FOV. Each telescope is equipped with the same filter system, denoted as \grizy\ \citep[as described by][]{Tonry2012_PS_Photo}. Images from both Pan-STARRS telescopes are processed with the Image Processing Pipeline \citep[IPP;][]{magnier2020a, waters2020}.

The individual exposure frames (called warps) are astrometrically and photometrically calibrated \citep{magnier2020c} and overlapping exposures are co-added with median clipping applied (to produce stacks). The Pan-STARRS1 $3 \pi$ reference sky images are subtracted from these frames \citep{waters2020} and photometry carried out on the resulting difference image \citep{magnier2020b}. These individual detections are ingested into the Pan-STARRS Transient Server database (now located in Oxford) and assimilated into distinct objects with a time variable history. A series of quality filters are applied using the IPP image attributes and known asteroids and variable stars are removed. The objects remaining are cross-matched with all catalogued galaxies, active galactic nuclei (AGN), cataclysmic variables (CVs) and historical transients \citep{Smith2020_ATLAS}, and simultaneously a machine learning algorithm is applied to image pixel stamps at each transient position \citep{2015MNRAS.449..451W,Smith2020_ATLAS}. This reduces the bogus detections to a manageable number for human scanning. For details on filtering and our human scanning process, specifically in the context of GW follow-up and faint transient identification, the reader is referred to \cite{Smartt2024_GW190425} and \cite{Fulton2025}.

We began observing the LVK \texttt{bilby.fits} skymap \citep{LVGCNdiscovery} on MJD 60994.24, 2.6 days post-merger (night of 2025 November 14; Hawai`i Standard Time). Our observations consisted of a series of dithered 45 and 200\,s exposures, reaching total exposure times per skycell of 90--720\,s (median exposure time of 180\,s) in the \ips\ and \wps bands. These images were combined into a single stacked image for each Pan-STARRS skycell \citep{Chambers2016arXiv_PanSTARRS1} on each night, and the standard processing as described above was immediately carried out. Typical 3.5$\sigma$ limiting magnitudes of m$_{\wps}$ $\simeq$ 17.8-22.9 were achieved on these nightly stacks, with the large range being due to varying sky conditions. We covered 1084 deg$^{2}$ (65\%) of the \texttt{bilby.multiorder.fits} 90\% sky localisation region with Pan-STARRS within 7 days post-merger. Our Pan-STARRS tiling of the S251112cm skymap is visually illustrated in Fig \ref{fig:skymap}.

\subsection{ATLAS}
\label{sec_sub:ATLAS_obs}
The ATLAS system consists of five units, situated at Haleakala and Mauna Loa (Hawai`i), El Sauce (Chile), Sutherland (South Africa), and Tenerife (Canary Islands) \citep{Tonry2018_ATLAS,Licandro2023}. Together, they tile the entire visible sky every $\sim$ 24\,hrs (weather-dependent). The telescopes survey the sky in cyan ($c$) and orange ($o$), and the Tenerife unit uses a ‘wide’ passband ($w$; CMOS response, a similar effective wavelength to the combined Sloan $g^\prime$ + $r^\prime$ bands). Typically, a pattern of 4 × 30\,s dithered exposures is acquired each night (roughly spaced logarithmically within a 1\,hr period). Our data processing pipeline follows standard practices; the images are photometrically and astrometrically calibrated using RefCat2 \citep{Tonry2018_ATLAS_REF}, before template subtraction is performed. As described by \cite{Smith2020_ATLAS}, quality cuts are applied, catalogued objects are cross-matched, and machine-learning algorithms are applied to the data to reject spurious detections and identify extragalactic transients \citep{Weston2024,Stevance2025}, and the surviving candidates are then scanned by eye.

ATLAS canvassed a substantial fraction of the \texttt{bilby.multiorder.fits} skymap within the 14 nights following the initial alert. These exposures consist of 30 s dithered exposures or deeper 110 s exposures. The ATLAS coverage within 14 days post-merger is illustrated in Fig. \ref{fig:skymap}. We sampled 1545 deg$^{2}$ (87\%) of the \texttt{bilby.multiorder.fits} 90\% sky localisation region with ATLAS between 0.2 and 14 days post-merger. Each ATLAS pointing reached typical 5$\sigma$ limiting magnitudes of $m_o \simeq$ 18.2--20.2 (110 s) and $m_w \simeq$ 16.8--20.1 (30 s).

\subsection{Vera C. Rubin Observatory}
\label{sec_sub:rubin_obs}
The NSF-DOE Vera C. Rubin Observatory operates the 8.4 m Simonyi Survey Telescope at Cerro Pach\'on, Chile, equipped with the Legacy Survey of Space and Time Camera (LSSTCam), which images a 9.6\,deg$^2$ FOV in one of six optical filters \citep{LSSTCam2025, Ivezic2019_rubin}. A standard 30\,s exposure has the potential to reach $5\sigma$ depths of $m_r \sim 24.5$ in good conditions, and a combination of wide field, depth, and rapid slew makes Rubin well suited to the deep wide area regime that is required for gravitational wave follow-up. At the time of S251112cm the observatory was in an early operations period of continued system optimisation prior to the beginning of sustained LSST observations.

Rubin observations for S251112cm were obtained through the ToO system, described in detail by \cite{MacBride2026}. Similarly to ATLAS and Pan-STARRS, an SSM merger did not automatically satisfy the automated ToO trigger criteria adopted for the LSST \citep{Andreoni2024}; rather, the newly formed Rubin ToO advisory board endorsed the alert for observation through the manual override mechanism. The first observations of the localisation region began on MJD 60994.076, 2.437 days post-merger. The observing strategy targeted the accessible portion of the 70\% localisation region in both the LSST $g$ and $i$ bands, using a single 30\,s visit per band. The reason for targeting the 70\% region rather than the 90\% region was to balance the follow-up against ongoing commissioning activities \citep{MacBride2026}. 

On the first night of observations, 849.1\,deg$^2$ (39.2\% of the 90\% localisation probability) was imaged by Rubin in $g$, and a smaller $i$-band region of 754.9\,deg$^2$ (33.5\%) was imaged due to the region setting early in the night from Chile. These first-night areas are as reported in GCN Circular 42707 \citep{Rubin_GCN10}, which did not account for chip and raft gaps in LSSTCam, and so slightly overestimate the true coverage. However, this unobserved area was recovered the following night. This pattern repeated for each subsequent night pair through 10 days post-merger with the exception of day 8 post-merger due to engineering. The visit pointings and $5\sigma$ limiting depths used in this work are from the Rubin pre-LSST visit database and are now available in the visit summary tables of Rubin Early Data Preview~2 (EDP2). Across the follow-up campaign a median $5\sigma$ depth of $m_g \simeq 24.5$ and $m_i \simeq 23.5$ was achieved, making this the deepest wide-area search for an electromagnetic counterpart to a GW event to date \citep{MacBride2026}. The Rubin coverage can be seen in Fig. \ref{fig:skymap}. 

As Rubin templates did not exist over most of the imaged area, templates from the Dark Energy Camera were constructed from co-added DES, DELVE, and DECADE imaging allowing difference imaging of the observations to be performed \citep{Abbott2021, DrlicaWagner2022, Anbajagane2025}. Candidate detections were filtered by rejecting negative and dipole subtractions, and requiring a minimum of three detections spanning multiple filters across multiple nights. Furthermore, likely contaminants were removed by cross-matching against the Wide-field Infrared Survey Explorer (WISE) catalogue \citep{Cutri2012} and the Million Quasars catalogue \citep{Flesch2023} to reject AGN, and against Gaia DR3 to reject stellar sources \citep{GaiaCollaboration2023}. 

The surviving sample was down-selected by an on-call team of Rubin Project members to fading candidates within a projected 50\,kpc of NED galaxies inside the GW volume, and fast rising candidates ($< -0.3$ mag/day) \citep{MacBride2026}. The resulting candidates were reported in GCN Circulars 42707 and 43257 \citep{Rubin_GCN10, Rubin_GCN16} and on the Transient Name Server. We derive our Rubin candidate list from these pre-screened events, rather than from Rubin alert-stream products. 

\begin{figure*}
    \centering
    \includegraphics[width=0.95\linewidth]{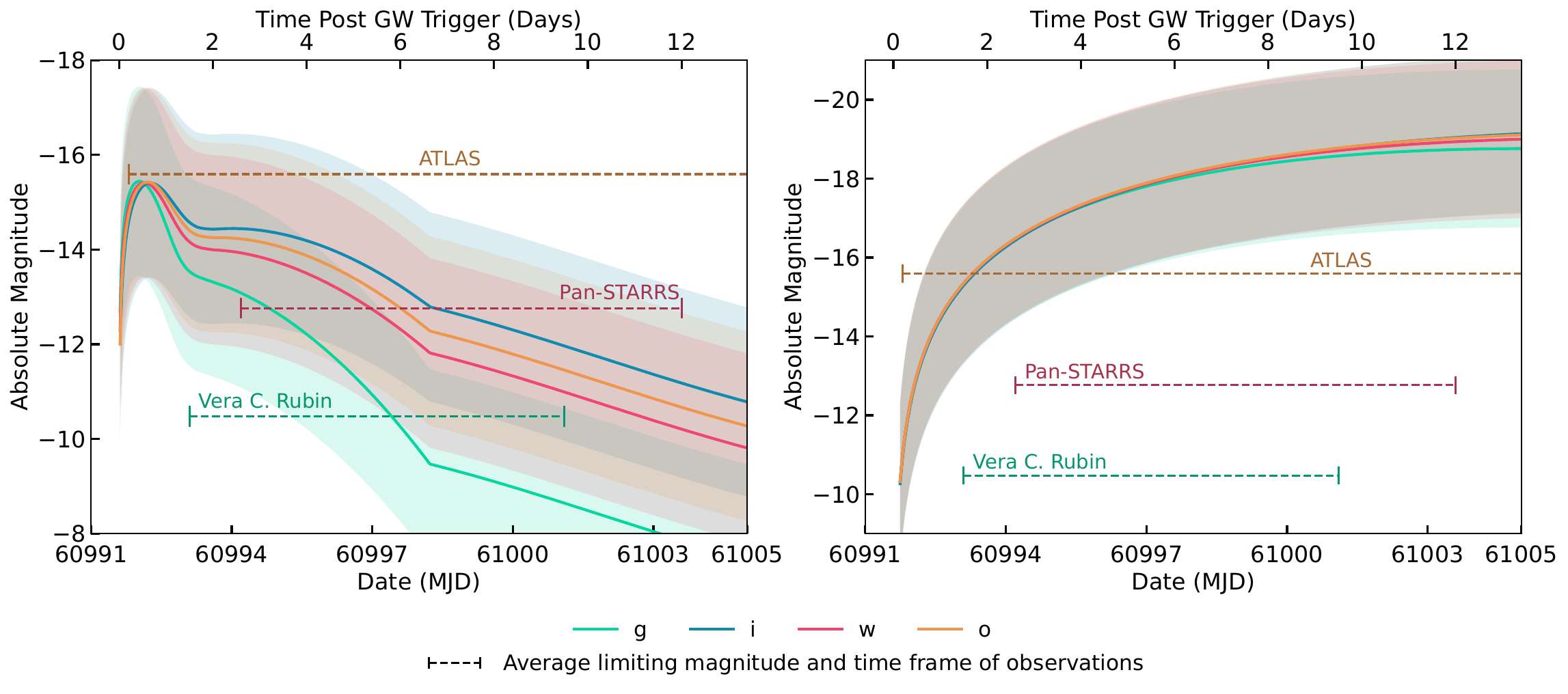}
    \caption{MOSFiT model light curves (absolute magnitude) for the AT~2017gfo-like kilonova (left) and the SN~1998bw-like Type Ic-BL supernova (right) in the LSST $g$ and $i$, Pan-STARRS $w$ and ATLAS $o$ bands, as a function of time since the GW trigger (top axis) and MJD (bottom axis). Shaded regions show the adopted model uncertainties of $\pm2$ mag (kilonova) and $\pm1$ mag (Type Ic-BL; Section \ref{secsub:MOSFiT_Models}). Horizontal lines mark the median limiting magnitude of each survey, converted to absolute magnitude at the LVK distance of 93\,Mpc, and span the time over which that survey observed the localisation region.}
    \label{fig:MOSFiT Models}
\end{figure*}

\subsection{VLT/FORS2 spectroscopy}

At the time of vetting, AT~2025alqe resided in a host galaxy with no spectroscopic or photometric redshift available through our Sherlock crossmatching, but with an apparent magnitude potentially consistent with the LVK distance posterior. We discuss this candidate in more detail in Section \ref{sec:final_candidates}.

To determine whether this candidate truly lies within the LVK distance, we obtained long-slit spectroscopy of its host with the FOcal Reducer and low dispersion Spectrograph 2 \citep[FORS2;][]{Appenzeller1998} on Unit Telescope 1 of the European Southern Observatory Very Large Telescope (VLT), through the Fast Track Channel (programme 117.2AX3; PI: A.~J.~Smith).

The observations were carried out in service mode on 2026 September 15 (MJD 61298.9), using a 1.3\,arcsec slit and the GRIS\_300V grism, giving coverage of $\sim$4500--8500\,\AA. The host was observed for a total of 2580\,s on source, split between two exposures. All VLT data were reduced with the ESO FORS2 pipeline in \texttt{EsoReflex} \citep{Freudling2013}, using standard calibrations.

\section{Counterpart models and candidate vetting} 
\label{sec:counterparts_and_vetting}
\subsection{MOSFiT models}
\label{secsub:MOSFiT_Models}
There are several possible EM counterparts for SSM mergers, all currently theoretical. Two isolated SSM NSs could produce a kilonova, similar to a classic BNS merger \citep{Metzger2017_kilonovae}, but few simulations exist in the SSM mass range to predict the expected ejecta mass. Alternatively, a SSM BNS merger could occur in a collapsar disc; this model associates the GW event with a Type Ic-BL supernova -- the only supernova class observationally associated with the collapsar scenario through their connection to long GRBs \citep{Chen_and_Metzger_SSM_Ic_BL, Metzger2024_SSM_Ic_BL}. Therefore we assume an AT~2017gfo-like kilonova and a SN~1998bw-like Type Ic-BL supernova as representative examples of the possible optical counterparts for our analysis, and generate template light curves to compare to candidates in the S251112cm localisation volume. AT~2017gfo is also representative of the known kilonova population: uniform MOSFiT modelling of the observed sample places it close to the sample median in most derived properties \citep{Rastinejad2025}.

We note that for both S251112cm and S250818k, a Type IIb supernova was discovered within the GW localisation region (SN~2025adtq and SN~2025ulz, respectively). \citet{Kasliwal2025_250818k} proposed that SN~2025ulz may be physically associated with S250818k, with the progenitor having undergone disc fragmentation and an SSM merger. However, the observational and theoretical connections between SNe IIb and SSM GW sources remain unproven \citep{Gillanders2025_250818k, ENGRAVE_250818k}. Moreover, given the large volumes enclosed by these localisation regions, and that SNe IIb comprise $\sim$11 per cent of core-collapse supernovae \citep{Graur_2017}, the serendipitous detection of an unrelated SN IIb is not unexpected; \citet{Hall2026_ZTF_S251112cm} quantify this chance-coincidence probability for S251112cm. We therefore do not include a SN IIb template in this study.

We construct our template for AT~2017gfo using publicly available multi-wavelength photometry compiled and described in detail by \citet{Villar2017_170817_Data}, compiled from published data \citep{Andreoni2017_170817, Arcavi_170817, Coulter2017_170817, cowperthwaite-2017, Diaz2017_170817, Drout2017_170817, Evans2017_170817, Hu2017_170817, valenti-2017, Kasliwal2017_270817,Lipunov2017_170817, pian-2017, Pozanenko2018_170817, Shappee2017_170817, Smartt2017, tanvir-2017, Troja2017_170817, Utsumi2017_170817}. This dataset incorporates ground-based optical and near-infrared observations alongside space-based Hubble Space Telescope and \textit{Swift}/UVOT data. For SN~1998bw, optical photometry in the $UBVRI$ bands was obtained from the Open Supernova Catalog \citep{Guillochon2017}\footnote{\url{https://github.com/astrocatalogs/sne-1990-1999/}}, which compiles observations from \citet{Galama1998}, \citet{McKenzie1999}, \citet{Sollerman2000}, \citet{Patat2001} and \citet{Sollerman2002}.

\begin{figure*}
    \centering
    \includegraphics[width=0.98\linewidth]{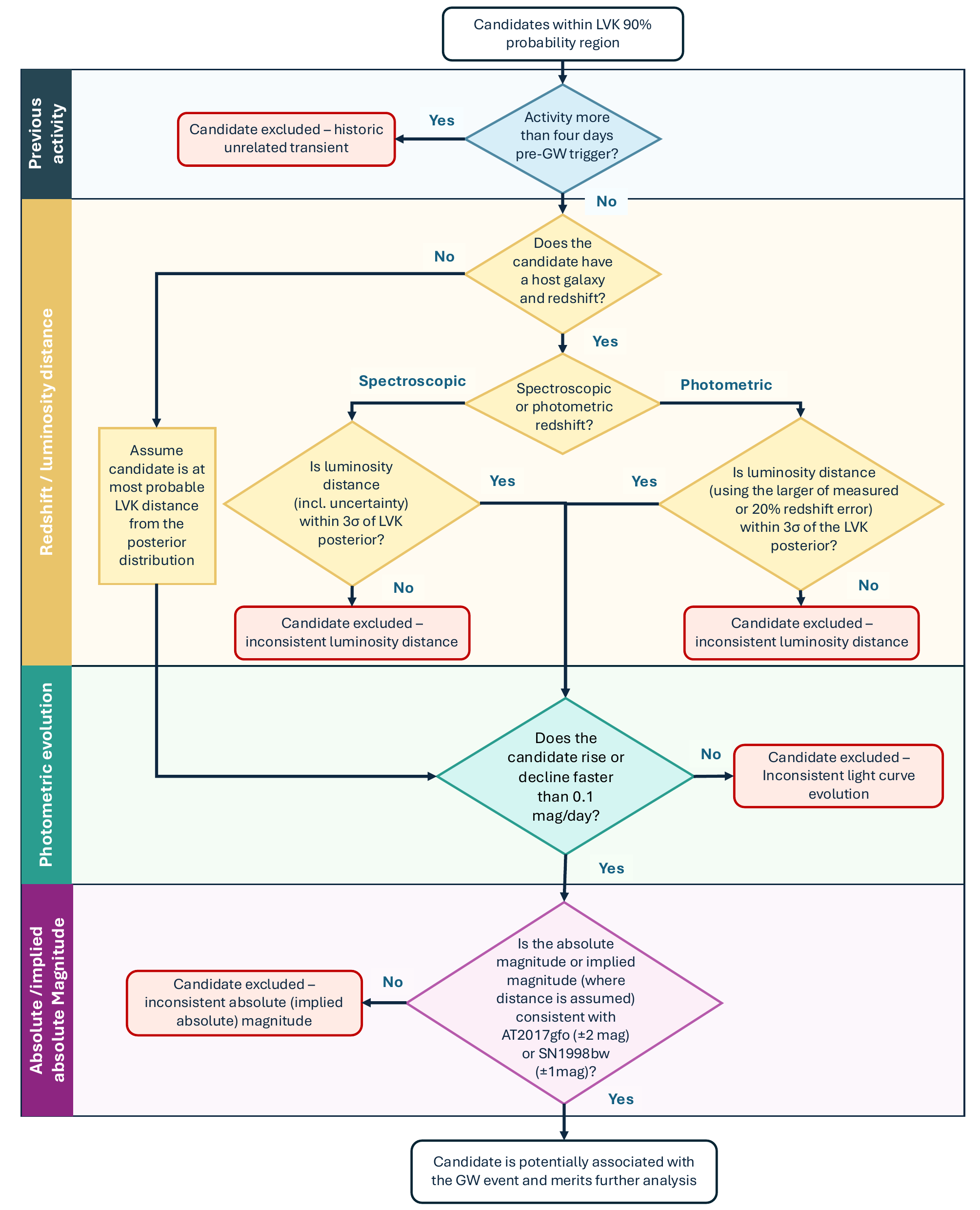}
    \caption{Flowchart of the four-stage candidate vetting procedure described in Section \ref{sec:candidate_vetting}. Candidates within the LVK 90\% localisation region are tested sequentially for (i) activity more than 4 days before the GW trigger, (ii) consistency of the host luminosity distance with the $3\sigma$ LVK distance posterior (adopting a $0.2z$ uncertainty for photometric redshifts, or assuming the most probable LVK distance where no redshift is available), (iii) photometric evolution rises/declines faster than 0.1\,mag/day at some point of the light curve, and (iv) the absolute (or implied absolute) magnitude is consistent with either model to within $\pm2$ or $\pm1$ mag of AT~2017gfo and SN~1998bw, respectively. Candidates passing all stages undergo the additional checks of Section \ref{sec_sub:candidates_further_analysed}.}
    \label{fig:flowchart}
\end{figure*}

The light curves of both transients were modelled using the Modular Open Source Fitter for Transients \citep[\texttt{MOSFiT}; ][]{Guillochon2018_MOSFiT}, employing dynamic nested sampling \citep[\texttt{dynesty}; ][]{Speagle2020_dynesty1} to explore the posterior parameter space. The fitting procedure was run until convergence was achieved, with the resulting best-fit posterior parameters for the \texttt{bns\_generative} model \citep{Nicholl2021_BNS_MOSFiT} applied to AT~2017gfo presented in Table \ref{tab:2017_mosfit_params}, and those for the modified \texttt{default} model \citep{Guillochon2018_MOSFiT, Nicholl2017_default_MOSFiT} applied to SN~1998bw presented in Table \ref{tab:1998bw_mosfit_params}. A minor modification was implemented for the SN~1998bw model: an additional UV suppression component was introduced following the methodology described in \cite{Gomez2022_UVSupression_MOSFiT}. The corresponding best-fit model light curves are shown in Fig. \ref{fig:MOSFiT Models}. 

Given the uncertain ejecta mass in the kilonova scenario and the diversity in SN Ic nickel mass, we allow for some flexibility in absolute magnitude when comparing to detected sources in the LVK volume. The range in magnitude for the AT~2017gfo-like model was taken to be $\pm$2 mag, based on the range of detections and upper limits for sGRB-associated kilonovae presented in \cite{Gompertz2018_KN_Diversity}.  Similarly, the Ic-BL error comes from peak scatter presented in \cite{Taddia2019_Ic_BL_Diversity}.  Because the scatter in peak magnitude is larger than the change in distance modulus over the localisation volume, allowing this spread in peak magnitude also accounts for the varying luminosity distance within the LVK distance posterior.

\begin{table}
 \begin{tabular*}{\columnwidth}{@{}l@{\hspace*{15pt}}l@{\hspace*{15pt}}l@{}}
  \hline
  Parameter & Description & Value\\
  \hline
  \texttt{Mchirp} & Chirp mass ($\rm M_{\odot}$) & $1.188$\\
  \texttt{q} & Mass ratio ($m_2/m_1$) & $0.92$\\
  \texttt{disc\_frac} & Disc wind ejection fraction & $0.15$\\
  \texttt{cos\_theta} & Cosine of viewing angle & $0.83$\\
  \texttt{shock\_frac} & Fraction of dynamical ejecta shocked by GRB & $0.5$\\
  \hline
 \end{tabular*}
  \caption{Best-fitting parameters of the MOSFiT \texttt{bns\_generative} model \citep{Nicholl2021_BNS_MOSFiT} fitted to the multi-wavelength photometry of AT~2017gfo compiled by \citet{Villar2017_170817_Data}. These parameters define the AT~2017gfo-like kilonova model light curves used for candidate vetting (Section \ref{sec:candidate_vetting}) and the exclusion analysis (Section \ref{sec:coverage_calc}).}
  \label{tab:2017_mosfit_params}
\end{table}

\begin{table}
 \begin{tabular*}{\columnwidth}{@{}l@{\hspace*{20pt}}l@{\hspace*{20pt}}l@{}}
  \hline
  Parameter name & Description & Value\\
  \hline
  \texttt{cutoff\_wavelength} & Blackbody cutoff wavelength (\AA) & $2650$\\
  \texttt{fnickel} & Nickel mass fraction & $0.21$\\
  \texttt{kappa} & Ejecta opacity (cm$^{2}$ g$^{-1}$) & $0.1$\\
  \texttt{kappagamma} & Gamma-ray opacity (cm$^{2}$ g$^{-1}$) & $0.1$\\
  \texttt{mejecta} & Ejecta mass ($\rm M_{\odot}$) & $2.6$\\
  \texttt{temperature} & Temperature floor (K) & $5850$\\
  \texttt{vejecta} & Ejecta velocity (km s$^{-1}$) & $24070$\\
  \hline
 \end{tabular*}
  \caption{Best-fitting parameters of the MOSFiT \texttt{default} (nickel-powered) model \citep{Guillochon2018_MOSFiT, Nicholl2017_default_MOSFiT}, modified to include UV suppression below the cutoff wavelength following \citet{Gomez2022_UVSupression_MOSFiT}, fitted to the $UBVRI$ photometry of SN~1998bw. These parameters define the SN~1998bw-like Type Ic-BL model light curves used for candidate vetting (Section \ref{sec:candidate_vetting}) and the volume exclusion analysis (Section \ref{sec:coverage_calc}).}
  \label{tab:1998bw_mosfit_params}
\end{table}

\subsection{Initial vetting}
\label{sec:candidate_vetting}

Candidates passing each survey's machine-learning real/bogus cuts were then visually inspected to reject any remaining image artefacts, such as poor subtractions, detector defects and moving objects. The four-stage vetting below was applied only to the candidates that survived this inspection.

To enable the rapid rejection of unlikely candidates in real time, we developed a four-stage vetting procedure that can assess candidates based on previous activity, redshift or luminosity distance, photometric evolution, and absolute magnitude (or implied absolute magnitude at the LVK distance, if redshift measurements are unavailable). This process is outlined below, and in Fig. \ref{fig:flowchart}. 

\begin{enumerate}
    \item \textbf{Previous activity} - No prior electromagnetic emission is expected for a kilonova \citep{Metzger2017_kilonovae}. However, with the uncertainty surrounding the proposed Type Ic-BL scenario \citep{Metzger2024_SSM_Ic_BL, Chen_and_Metzger_SSM_Ic_BL, Hall2026_ZTF_S251112cm}, a 4 day buffer was permitted for candidate selection. This accounts for the presently unconstrained timescales between the initial Type Ic-BL explosion, the formation of SSM NSs within the accretion disc, and their subsequent merger(s) and GW signal. 
    \item \textbf{Redshift/luminosity distance} - Candidates were cross-matched against their likely host galaxies using our \texttt{Sherlock} intelligent cross-matching algorithm \citep{Smith2020_ATLAS}. Hosts with catalogued host redshifts were compared against the LVK distance posterior. For spectroscopic measurements, the luminosity distance, including measurement error, had to overlap the $3\sigma$ range of the LVK distance constraint. Photometric redshift measurements were treated the same way; a candidate was retained if its host redshift overlapped the 3$\sigma$ LVK distance range using either its catalogued uncertainty or an assumed 20\% error, motivated by the photometric redshift uncertainties reported for Sloan Digital Sky Survey (SDSS) galaxies by \citet{Beck2016_SDSS_z_Errors}. Photometric redshifts based on Pan-STARRS data were not available within \texttt{Sherlock} and so are discussed separately in Section \ref{sub_sec:PS_photozs}. For S251112cm, the LVK distance was $93 \pm 27$\,Mpc, corresponding to an acceptable distance range of 12--174\,Mpc. In order to be conservative at this stage in our vetting, candidates for which no redshift estimates were available in \textsc{Sherlock} were assumed to lie at the LVK-inferred distance of 93\,Mpc.
    \item \textbf{Photometric evolution} - For each candidate we computed the rate of change in magnitude between every pair of post-trigger detections in the same band. Candidates were retained if any of these rates exceeded 0.1\,mag\,d$^{-1}$, either fading (kilonova-like) or rising (consistent with a young supernova), and excluded otherwise. This is a deliberately inclusive threshold, well below the decline of our kilonova model ($\gtrsim 0.3$\,mag\,d$^{-1}$ at 4--5\,d) and the early rise of the SN~1998bw model.
    \item \textbf{Absolute/implied absolute magnitude} - The distance modulus for each remaining candidate was calculated using either the host-derived luminosity distance where possible or the assumed LVK distance. Absolute magnitudes were then compared against the model absolute magnitudes at the corresponding post-trigger epoch. Uncertainties in the models were taken to be $\pm2$\,mag and $\pm1$\,mag for the kilonova and Type Ic-BL model, respectively. This range in magnitude is justified in Section \ref{secsub:MOSFiT_Models}. Candidates were retained only where both the photometric evolution and absolute magnitude were mutually consistent with the same physical model. For example, a candidate exhibiting a rising light curve consistent with the Type Ic-BL model, but possessing an absolute magnitude only consistent with the kilonova model at that epoch, was rejected owing to the discrepancy between its luminosity and temporal evolution.
\end{enumerate}

\subsection{Manual inspection}
\label{sec_sub:candidates_further_analysed}
After the four-stage vetting, a number of candidates remained. These candidates remained generally due to lack of observations, with fewer than five nights of observations often exhibiting potentially compatible light curve evolution at a similar expected magnitude with either of the AT~2017gfo-like kilonova or SN~1998bw-like Type Ic-BL models in host galaxies lacking redshift or distance measurements.

For each remaining candidate, we visually compared the absolute magnitude to the relevant light curve model. Furthermore, the candidate host galaxies were visually inspected and their implied absolute magnitudes at 93\,Mpc compared against the distribution of host galaxy magnitudes in the ATLAS100 volume-limited catalogue \citep{Srivastav_2026}.

\section{Results}
\label{sec:results}
\subsection{Candidate vetting results}
\label{secsub:survey_comparison}
\begin{figure*}
    \centering
    \includegraphics[width=0.95\linewidth]{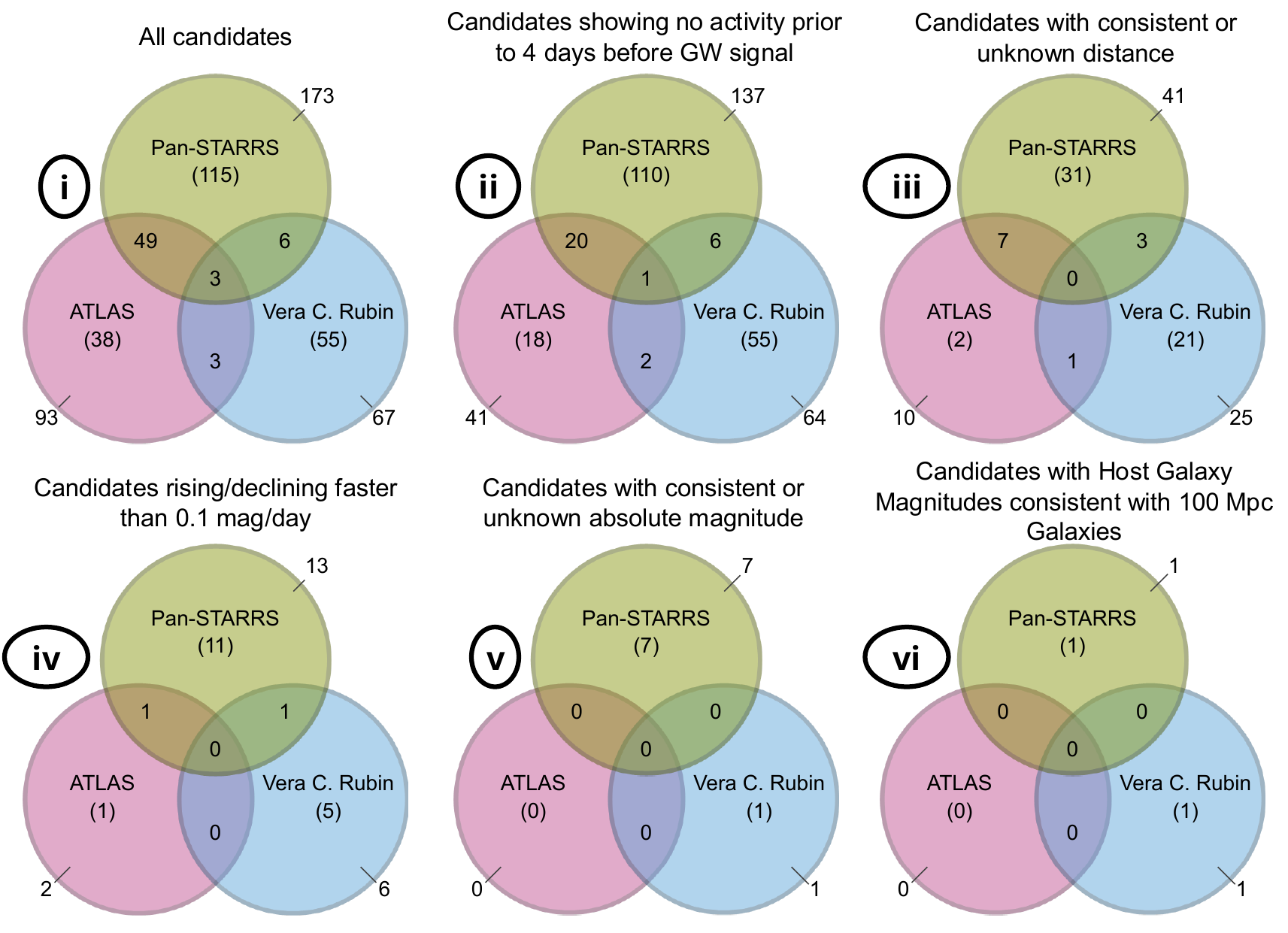}
    \caption{Number of candidates from ATLAS (purple), Pan-STARRS (green) and Rubin (blue) remaining after each vetting stage. Panel (i) shows all 269 unique candidates; panels (ii)--(v) correspond to stages (i)-(iv) of Fig. \ref{fig:flowchart}; panel 6 shows candidates whose host galaxies overlap the ATLAS100 host magnitude distribution (Section \ref{sec_sub:candidates_further_analysed}). The eight candidates that remained up to stage (v) are shown alongside their hosts in Fig. \ref{fig:edgecase_plots}, in particular AT~2025alqe and AT~2025aimy which are the two candidates remaining in stage (vi).}
    \label{fig:venndiagram}
\end{figure*}

After machine learning rejection of artefacts and human scanning, ATLAS found 93 reliable extragalactic candidates, Pan-STARRS 173, and Rubin 67. However, these Rubin candidates were pre-filtered as outlined in Section~\ref{sec_sub:rubin_obs}, retaining only fading sources near catalogued galaxies within the GW volume and fast risers; the Rubin sample is therefore biased towards stronger candidates than those from ATLAS and Pan-STARRS, and its stage-by-stage survival fractions are not directly comparable. Of these, 61 are reported by multiple surveys, giving a total of 269 unique candidates (see stage (i), Fig. \ref{fig:venndiagram}). After removing all candidates showing activity earlier than four days pre-GW, the number (fraction) of remaining candidates in each survey was: ATLAS - 41 (44\%), Pan-STARRS - 137 (79\%), and Rubin - 64 (96\%), leaving 212 unique candidates (See stage (ii), Fig. \ref{fig:venndiagram}). The near-daily sky coverage of ATLAS ensured that a large fraction of pre-existing transients could be removed by this simple cut. Rubin is least impacted by this previous activity cut, as fewer of its candidates had constraints on previous activity (the full Wide-Fast-Deep survey was not yet running). 

Removing candidates with catalogued redshift measurements in disagreement with the 3$\sigma$ LVK distance posterior left ATLAS with 10 (11\%), Pan-STARRS with 41 (24\%), and Rubin with 25 (37\%), leaving 65 unique candidates (See stage (iii), Fig. \ref{fig:venndiagram}). Already 89\% of ATLAS candidates could be rejected; its shallower depth means that most of the transients it detects are in nearby hosts more likely to have a measured redshift. 

The cut on light curve morphology removes most of the remaining transients. After removing candidates that do not rise or decline faster than 0.1\,mag/day at any stage of their evolution, the number of candidates remaining in ATLAS was 2 (2\%), Pan-STARRS 13 (8\%) and Rubin 6 (9\%) (See stage (iv), Fig. \ref{fig:venndiagram}). The excluded objects have flat or slowly fading light curves consistent with normal supernovae around peak brightness; the number of unique remaining candidates was reduced to 19.

Finally, after calculating absolute magnitudes -- either at the catalogued distance if known, or at 93\,Mpc if not -- and comparing against the models with their uncertainties, only eight candidates remained. ATLAS -- 0, Pan-STARRS -- 7 (4\%), and Rubin -- 1 (1\%) (See stage (v), Fig. \ref{fig:venndiagram}). The additional manual checks discussed in Section \ref{sec_sub:candidates_further_analysed} were then employed on these candidates. 

\subsection{Final inspection of remaining candidates}
\label{additional_vetting_results}
The remaining eight candidates that are either consistent or lack sufficient data to exclude them using the vetting stages shown in Fig. \ref{fig:flowchart} are shown plotted with their light curves and host galaxies in Fig. \ref{fig:edgecase_plots}. All remaining candidates have shown some form of decline in their light curve at the same post-merger time epoch as AT~2017gfo, and hence are plotted alongside the AT~2017gfo-like kilonova model. 

Three of these candidates, AT~2025alph, AT~2025adno and AT~2025aimy, fade markedly more slowly than a kilonova. They decline at rates of $\sim 0.1$\,mag/day between 2--10 days post-merger, which is enough to pass our initial light curve cut. They also have absolute magnitudes in the kilonova range if placed at the distance of S251112cm. However, their evolution over several days is clearly slower than the rapid decline expected of a kilonova, based on visual inspection of the light curves. The light curve of AT~2025aimy is particularly well observed, which we discuss further in Section \ref{sec:final_candidates}.

\begin{figure*}
     \centering
     \includegraphics[width=0.7\linewidth]{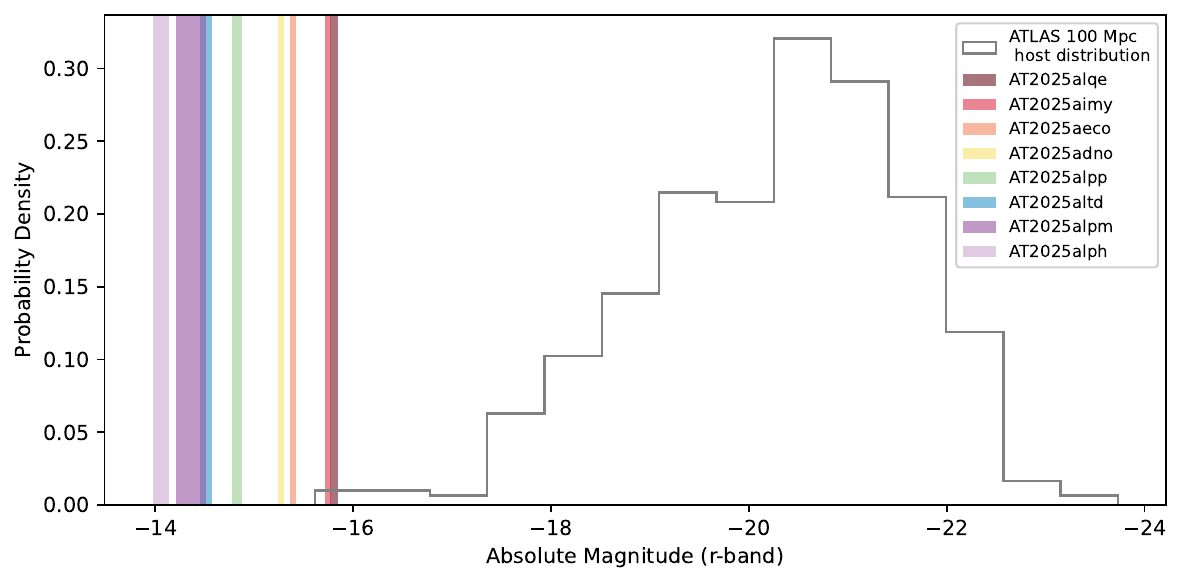}
     \caption{Implied $r$-band absolute magnitudes of the host galaxies of the eight remaining candidates (vertical bands), compared with the normalised host galaxy absolute magnitude distribution of transients in the volume-limited ATLAS100 sample \citep{Srivastav_2026}. Host magnitudes are computed assuming the LVK distance of 93\,Mpc, except for AT~2025aimy, which uses its spectroscopic distance of 140\,Mpc. For candidates with more than one plausible host, only the brightest is shown. Six of the eight hosts are fainter than every galaxy in the ATLAS100 sample, and the remaining two lie below its first percentile (Table \ref{tab:host_mag_dist}).}
     \label{fig:abs_mag_hist}
\end{figure*}

Figure \ref{fig:abs_mag_hist} shows the implied absolute magnitude of each of the candidates' host galaxies were they located at a luminosity distance of 93\,Mpc, with the exception of AT~2025aimy which has a spectroscopic redshift placing it at 140\,Mpc. For comparison, the absolute magnitudes of transient host galaxies from the ATLAS100 survey are also shown. It is clear that six of the eight host galaxies are fainter than any in the ATLAS100 distribution, and only the host galaxies of AT~2025alqe and AT~2025aimy overlap the general host population. However, as shown in Table \ref{tab:host_mag_dist}, these two hosts still lie below the first percentile of the faintest transient hosts in the local Universe. 

Placing these hosts at the distance of S251112cm therefore requires them to be intrinsically fainter than almost every galaxy known to host a nearby transient. The more natural interpretation is that most are background galaxies, whose apparent faintness reflects a greater distance rather than a low luminosity, so that their transients are unrelated to the GW event. AT~2025aimy is the one candidate whose host distance is measured directly and is consistent with the GW localisation volume. Even so, its host's low luminosity makes it an unlikely site for a compact-object merger, particularly in the isolated binary channel, where the merger rate is expected to scale with stellar mass.

\subsection{Additional constraints on host galaxy redshift}
\label{sub_sec:PS_photozs}
\begin{table}
    \centering
    \renewcommand{\arraystretch}{1.4}
    \begin{tabular}{lccc}
        \hline
        Host & $z_{\rm phot}$ & $\sigma_{z}$ & $D_\mathrm{L}$ (Mpc) \\
        \hline
        AT~2025alpm host 1 & -- & -- & -- \\
        AT~2025alpm host 2 & 0.49 & 0.19 & $2792^{+1332}_{-1202}$ \\
        AT~2025alph host & 0.30 & 0.18 & $1538^{+1142}_{-990}$ \\
        AT~2025altd host & 0.44 & 0.14 & $2423^{+983}_{-902}$ \\
        AT~2025alqe host & 0.14 & 0.04 & $669^{+236}_{-224}$ \\
        AT~2025alpp host & 0.18 & 0.12 & $851^{+697}_{-615}$ \\
        AT~2025adno host 1 & -- & -- & -- \\
        AT~2025adno host 2 & -- & -- & -- \\
        AT~2025adno host 3 & 0.43 & 0.13 & $2388^{+860}_{-796}$ \\
        AT~2025aeco host 1 & 0.15 & 0.05 & $730^{+269}_{-255}$ \\
        AT~2025aeco host 2 & 0.61 & 0.06 & $3596^{+410}_{-400}$ \\
        \hline
    \end{tabular}
    \caption{Photometric redshift measurements from Pan-STARRS1 catalogue for the host galaxies of the kilonova candidates mentioned in Section \ref{additional_vetting_results}. Photometric redshifts were retrieved from the catalogue of \citet{Tarrio2020}. In instances where the same host returned more than one redshift, we adopted the one that yielded the smallest luminosity distance to be conservative, and the quoted luminosity distance bounds correspond to $z_{\rm phot} \pm \sigma_{z}$.}
    \label{tab:photoz}
\end{table}

The above host luminosity analysis suggests that all but AT\,2025aimy are likely to be in galaxies beyond the LVK localisation distance. However, this assumes they are drawn from the same galaxy population as the ATLAS100 sample (likely, but not certain). We therefore retrieved additional photometric redshift measurements for the remaining candidate host galaxies from the Pan-STARRS1 catalogue of \citet{Tarrio2020}. As noted in Section \ref{sec:candidate_vetting}, these redshifts were not available in our initial \texttt{Sherlock}-based cross-matching.

The catalogue applies the local linear regression method of \cite{Beck2016_SDSS_z_Errors} in the five-dimensional space of $g-r$, $r-i$, $i-z$ and $z-y$ Kron colours plus the $r$-band Kron magnitude, and is restricted to just primary detections for which at least four of the five are available. Of the eight candidates, there were 11 possible host galaxies (i.e.~some transients could be plausibly associated with more than one host) that did not have a redshift measurement (spectroscopic or photometric) in \texttt{Sherlock}. Of these eleven, eight had photometric redshifts in this catalogue. However, the most likely host of AT~2025alpm (denoted host 1 in Fig. \ref{fig:edgecase_plots}) did not have a redshift, while the other host candidate (host 2) did. For AT~2025adno, which has three possible hosts, only host 3 had a redshift measurement. See Table \ref{tab:photoz}.

The three host galaxies not included in this catalogue (AT~2025alpm host 1 and AT~2025adno hosts 1 and 2) are not incidental, as the feature availability requirement to estimate the redshift systematically excludes such faint host galaxies that are not detected in one or more bands. However, as they are among the faintest galaxies in the sample, an association with any of these hosts most likely indicates a distance beyond the GW volume.

For those with Pan-STARRS photometric redshifts, the corresponding luminosity distances, including $1 \sigma_z$ redshift errors, place all of the host galaxies outside of the $3\sigma$ upper bound of the LVK distance posterior for S251112cm. Furthermore, this is consistent with our earlier conclusion that these remaining transients are associated with distant host galaxies and are unrelated to S251112cm.

\subsection{AT~2025aimy and AT~2025alqe}
\label{sec:final_candidates}

\begin{figure*}
     \centering
     \includegraphics[width=0.98\linewidth]{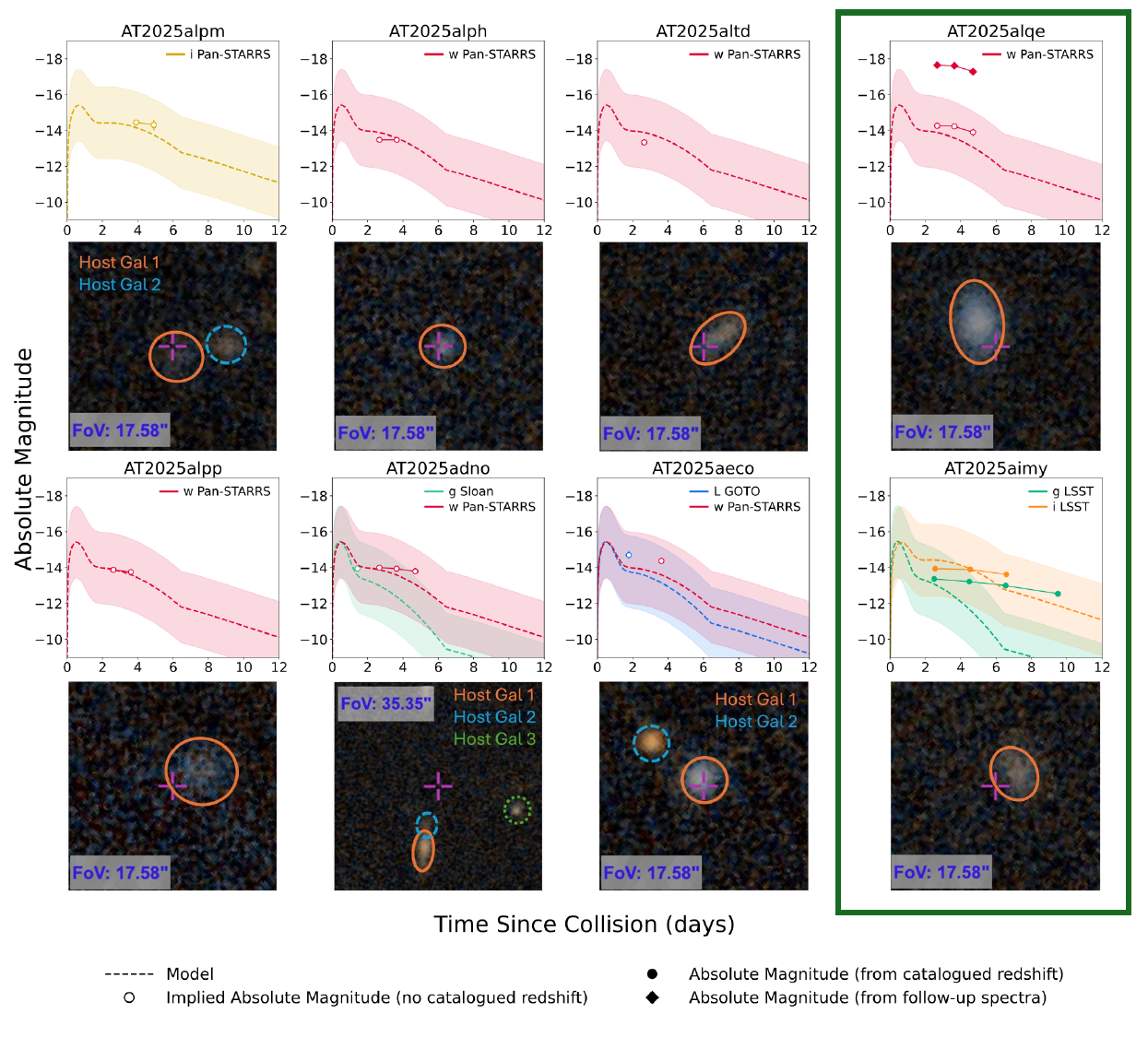}
     \caption{Light curves of the eight candidates surviving the four-stage vetting, compared with the AT~2017gfo-like kilonova model (dashed lines; shaded regions show the $\pm2$ mag uncertainty, Section \ref{secsub:MOSFiT_Models}) as a function of time since the GW trigger. Filled markers denote absolute magnitudes derived from a spectroscopic redshift; open markers denote implied absolute magnitudes assuming the LVK distance of 93\,Mpc. For AT2025alqe, we show both open symbols representing the light curve used in our real-time vetting and filled symbols showing the true luminosity after we obtained a host spectrum.  Additional photometry was taken from the TNS for AT~2025adno (GW-MMADS; \citealt{Hall_2025_TNS}) and AT~2025aeco (GOTO; \citealt{ONeill_2025_TNS}). Below each light curve is a PS1 colour image of the field (field of view indicated) \citep{Chambers2016arXiv_PanSTARRS1, Boch2019_aladdin}, with the transient position marked by the purple cross and candidate host galaxies circled and labelled where more than one is plausible. These candidates host galaxy magnitudes can be seen compared against the ATLAS100 host galaxy magnitude distribution in Fig. \ref{fig:abs_mag_hist}. The green boxes highlight AT\,2025alqe and AT\,2025aimy, the two candidates of particular interest discussed further in Section~\ref{sec:final_candidates}.}
     \label{fig:edgecase_plots}
\end{figure*}

Following the host galaxy analysis, only two possible candidate counterparts remain:

\textit{AT~2025aimy} - AT~2025aimy showed a decline rate of 0.14\,mag/day in the LSST $i$ band between MJD 60996 and 60999. However, the LSST $g$-band observations show a decline of only 0.83\,mag between 3 and 10 days post merger, compared with 4.38\,mag for AT~2017gfo over the same interval. As noted in Section \ref{additional_vetting_results}, this evolution is far slower than expected for an AT~2017gfo-like kilonova. This is clear from Fig. \ref{fig:edgecase_plots}, and given the excellent $g$-band coverage from Rubin, we no longer consider AT~2025aimy a viable candidate.

\begin{figure}
    \centering
    \includegraphics[width=0.98\linewidth]{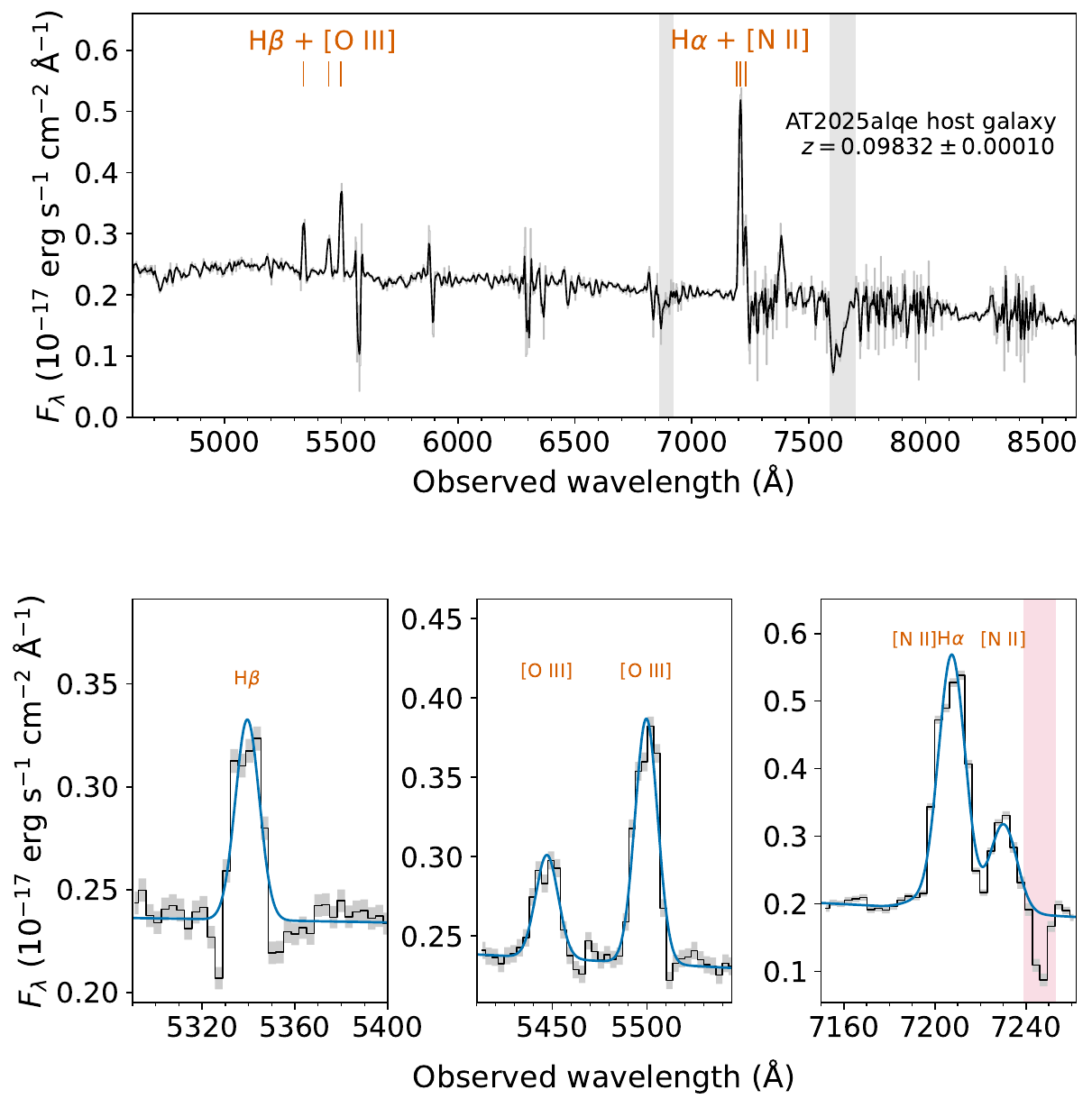}
    \caption{VLT/FORS2 spectrum of the host galaxy of AT~2025alqe. \textit{Top:} the full spectrum (grey) and the same spectrum smoothed with a three-pixel running mean (black). Grey shaded bands mark the telluric O$_2$ B-band (6860--6920\,\AA) and A-band (7590--7700\,\AA) absorption. Orange ticks show the observed positions of the fitted emission lines at $z = 0.09832 \pm 0.00010$, the inverse-variance weighted mean of the redshifts fitted to the three line groups. Isolated lines from the top plot that highlight the following emission lines: H$\beta$ (\textit{bottom left}), the [O\,\textsc{iii}] doublet (\textit{bottom middle}), and H$\alpha$ + [N\,\textsc{ii}] doublet (\textit{bottom right}). The black line is the observed flux, the grey shading is the $1\sigma$ uncertainty, and the blue line is the best fit: a linear continuum plus Gaussian profiles. The pink band in the H$\alpha$ + [N\,\textsc{ii}] panel marks a sky-subtraction residual that was excluded from the fit. The redshift corresponds to $D_\mathrm{L} \approx 452$\,Mpc, well beyond the $3\sigma$ upper limit of 174\,Mpc for S251112cm.}
    \label{fig:alqe_spectrum}
\end{figure}

\textit{AT~2025alqe} - AT~2025alqe was detected in only three epochs of Pan-STARRS $w$-band imaging, over which it declined at $\sim$0.2\,mag/day. Given this decline -- albeit with limited photometric coverage -- we could not rule it out, resulting in AT~2025alqe remaining a viable candidate. If placed at the LVK distance of 93\,Mpc, its host was also just luminous enough to overlap with the ATLAS100 galaxy distribution (Section~\ref{additional_vetting_results}). We therefore obtained a spectrum of the host with VLT/FORS2 (programme 117.2AX3). The spectrum shows narrow H$\alpha$ and H$\beta$ emission, along with the [O\,{\sc iii}] and [N\,{\sc ii}] doublets (Fig.~\ref{fig:alqe_spectrum}). Fitting these with Gaussian profiles gives a redshift of $z = 0.09832 \pm 0.00010$, corresponding to a luminosity distance of $D_\text{L} \approx 452$\,Mpc. This lies far beyond the 3$\sigma$ upper limit of 174\,Mpc on the distance to S251112cm ($93 \pm 27$\,Mpc), placing the host $\sim$13$\sigma$ from the GW distance. This is consistent within 1$\sigma$ with the Pan-STARRS1 photometric redshift of the host ($z_{\rm phot} = 0.14 \pm 0.04$; Table~\ref{tab:photoz}, \citealt{Tarrio2020}), which was not available in Sherlock at the time of vetting (Section~\ref{sub_sec:PS_photozs}). At this distance, the absolute magnitude of AT~2025alqe is $-17.6$\,mag, significantly brighter than our kilonova model range at the same phase. We therefore rule out AT~2025alqe as the counterpart.

\subsection{Candidates of interest from other surveys}

Other teams have applied their own searches and approaches to ranking candidates in the S251112cm volume \citep{Hall2026_ZTF_S251112cm, Vieira2026_TROVE}. We discuss here the candidates flagged as potentially interesting by other studies, and why they were not considered to be of interest in terms of our vetting framework. 

\textit{AT~2025adin} - AT~2025adin is a candidate of interest as a possible kilonova candidate in the TROVE vetting framework \citep{Vieira2026_TROVE}, as it had a spectroscopic redshift placing it at 87\,Mpc. This places it within our tolerance of the LVK distance posterior, and it showed no evidence of previous activity. However, after looking at the light curve evolution from Rubin, the steepest decline was found to be 0.06 mag/day at 4-5 days post-merger, which we deemed inconsistent with the 0.33 mag/day rate at the same epoch in our AT~2017gfo-like kilonova model. 

\textit{AT~2025admd} - This candidate was also noted by \cite{Vieira2026_TROVE} due to it receiving a high 2D localisation score (in a high probability region of the projected skymap). However, with our constraints on luminosity distance excluding candidates outside of the 3$\sigma$ LVK distance posterior range, and this candidate being at photometric $z=0.11 \pm 0.05$ or 510$^{+256}_{-241}$\,Mpc \citep[Legacy Survey DR10; ][]{Zhou2023}, it was excluded at the luminosity distance stage of our vetting.

\textit{AT~2025ailp} - this candidate attracted attention with the addition of late-time photometry in the g-band showing a faster decline of 0.27 mag/day \citep{Vieira2026_TROVE}. As with AT~2025admd, we associated this candidate with a host with photometric redshift of $z = 0.16 \pm 0.02$ \citep[Legacy Survey DR9; ][]{Zhou2021}, which was found to be in a host at a luminosity distance of 765$^{+106}_{-104}$\,Mpc. 

\textit{AT~2025ails} - AT~2025ails was noted to be of interest by \citet{Vieira2026_TROVE} due to a rebrightening in its light curve. In our analysis, we found that AT~2025ails was associated with a host galaxy at a photometric redshift of $z=0.41 \pm 0.03$ (2237$^{+195}_{-192}$\,Mpc), and hence ruled this candidate out at the distance cuts placed at Stage (iii) of the vetting process \citep[Legacy Survey DR10; ][]{Zhou2023}. 

\textit{SN~2025adtq} - This event was a Type IIb supernova first discovered by Pan-STARRS, and noted to be coincident with S251112cm by \citet{Hall2026_ZTF_S251112cm}. This was flagged as potentially noteworthy due to the occurrence of a coincident Type IIb supernova also in S250818k \citep{Gillanders2025_250818k, Kasliwal2025_250818k, ENGRAVE_250818k}. \citet{Kasliwal2025_250818k} suggested a possible collapsar disc fragmentation scenario \citep{Metzger2024_SSM_Ic_BL} could link the supernova and GW signal, and result in a kilonova-in-supernova counterpart. In our analysis, we assume that this scenario requires a rapidly rotating, massive progenitor and likely the formation of a successful central engine, for which an accompanying Gamma Ray Burst (GRB) would be expected. As spectroscopically confirmed GRB-SN associations to date have been exclusively Type Ic-BL events \citep{Cano2017, Finneran2025}, this was the supernova template we used as a comparison case in our analysis. The first ATLAS observation showing a consistent rise in flux above the baseline was on MJD 60995.6, and the host is at a luminosity distance of 146\,Mpc. This led to SN~2025adtq passing Stages i-iv. However, SN~2025adtq did not pass Stage v, as its absolute magnitude at the corresponding time post-explosion was outside the region of uncertainty in our SN~1998bw-like SN Ic-BL model. While we do not speculate further on the claimed link between SNe IIb and SSM GW sources, future comprehensive searches of SSM skymaps should reveal whether there is a connection \citep{Gillanders2025_250818k,Kasliwal2025_250818k, ENGRAVE_250818k,Hall2026_ZTF_S251112cm}. 

Likewise, \citet{Paek2026} report no convincing counterpart.


\begin{table*}
    \centering
    \begin{tabular}{|c|c|c|c|c|c|}
    \hline
        Transient host & \shortstack[c]{Apparent \\ r-band magnitude} & \shortstack[c]{Absolute/implied \\ absolute magnitude} & \shortstack[c]{Host galaxy \\ magnitude error} & Distance / Mpc & \shortstack[c]{Percentile of ATLAS100\\ host magnitude distribution / \%} \\ \hline
        AT~2025alpm host  1 & 20.57 & $-14.36$ & 0.09 & 93 (assumed) & 0 \\ 
        AT~2025alpm host  2 & 21.22 & $-13.71$ & 0.08 & 93 (assumed) & 0 \\
        AT~2025alph host & 20.88 & $-14.05$ & 0.08 & 93 (assumed) & 0 \\ 
        AT~2025altd host & 20.42 & $-14.51$ & 0.06 & 93 (assumed) & 0 \\ 
        \textbf{AT~2025alqe host} & \textbf{19.13} & \textbf{$-$15.80} & \textbf{0.04} & \textbf{93 (assumed)} & \textbf{0.4} \\ 
        AT~2025alpp host & 20.11 & $-14.82$ & 0.05 & 93 (assumed) & 0 \\ 
        AT~2025adno host 1 & 19.66 & $-15.27$ & 0.03 & 93 (assumed) & 0 \\ 
        AT~2025adno host 2 & 21.61 & $-13.32$ & 0.26 & 93 (assumed) & 0 \\ 
        AT~2025adno host 3 & 20.90 & $-14.03$ & 0.06 & 93 (assumed) & 0 \\ 
        AT~2025aeco host 1 & 19.54 & $-15.39$ & 0.02 & 93 (assumed) & 0 \\ 
        AT~2025aeco host 2 & 20.81 & $-14.12$ & 0.06 & 93 (assumed) & 0 \\ 
        \textbf{AT~2025aimy host} & \textbf{19.95} & \textbf{$-$15.97} & \textbf{0.03} & \textbf{140} & \textbf{0.2} \\ \hline
    \end{tabular}
    \caption{Host galaxy $r$-band photometry for the eight remaining candidates. Absolute magnitudes are implied assuming a distance of 93\,Mpc unless a spectroscopic redshift is available (AT~2025aimy). The final column gives the percentile of each host within the ATLAS100 host galaxy absolute magnitude distribution \citep{Srivastav_2026}; a value of 0 indicates the host is fainter than every galaxy in that sample. Bold: the two candidates whose hosts overlap the ATLAS100 distribution.}
    \label{tab:host_mag_dist}
\end{table*}

\subsection{Quantifying the excluded localisation volume}
\label{sec:coverage_calc}
We quantify constraints placed on our non-detections using the method outlined by \citet{Gillanders2025_250818k}, to which we refer the reader for a more detailed explanation. In brief, we take the $5\sigma$ limiting magnitude for each image, and the absolute magnitude of the light curve models (Section \ref{secsub:MOSFiT_Models}) at the epoch of the observation (with time zero corresponding to the GW signal), with Galactic extinction applied for that line of sight, and determine the luminosity distance for which the model would fall below the limiting magnitude, this represents the maximum distance of a possible detection. Unlike the vetting in Section \ref{secsub:survey_comparison}, which adopts a sky-averaged distance, here we use the three-dimensional LVK skymap which supplies the distance posterior for each line of sight. With this we then integrate the posterior across all distances closer than that limit. This method returns the fraction of the distance posterior for which the observation can exclude a given model. A value of 1 indicates that the observation was deep enough to have observed a transient of the same characteristics as the model anywhere within the distance posterior of the GW signal in that direction. Where multiple observations of the same pointing were taken over time, we retain only the most constraining (for kilonovae this is usually the earliest observation, when the model is brightest, if observing conditions are similar). We report both the fraction of the observed area with observations deep enough to detect the model transient at some epoch, and the fraction of the complete 90\% localisation volume over which each survey was sensitive. 

This analysis is performed for ATLAS, Pan-STARRS and the Rubin observations individually in Sections \ref{sec_sub:ATLAS_Coverage}, \ref{sec_sub:Pan-STARRS_Coverage}, and \ref{sec_sub:Rubin_Coverage}, and their combined coverage is discussed in Section \ref{sec_sub:Collective_Coverage}. Individual results for each survey can be seen in Table \ref{tab:volume_excluded}. 

\begin{table*}
    \centering
    \setlength{\tabcolsep}{6.5pt}
    \begin{tabular}{l|c|cc|cc}
        \hline
         & & \multicolumn{2}{c|}{AT~2017gfo-like} & \multicolumn{2}{c}{SN~1998bw-like} \\
         & Total observing area & Fraction of & Fraction of 90\% & Fraction of & Fraction of 90\% \\
         & / deg$^{2}$ & observed volume & localisation volume & observed volume & localisation volume \\
        Survey & (\% of 90\% area) & excluded / \% & excluded / \% & excluded / \% & excluded / \% \\
        \hline
        ATLAS      & 1545 (87\%) & 14  & 12 & 81  & 70 \\
        Pan-STARRS & 1084 (65\%) & 99  & 64 & 100 & 65 \\
        Rubin      &  849 (36\%) & 100 & 36 & 100 & 36 \\
        \hline
    \end{tabular}
    \caption{Sky coverage of the S251112cm 90\% localisation region (1681.1~deg$^{2}$) and the fraction of the LVK localisation volume within which ATLAS, Pan-STARRS and Rubin exclude an AT~2017gfo-like kilonova and a SN~1998bw-like Type Ic-BL supernova. The first column gives the total area observed by each survey, with the percentage of the 90\% localisation probability covered by these observations in brackets (note that these are not necessarily equal to area divided by 1681.1\,deg$^2$, as not all pointings lie entirely within the 90\% region). For each model, the first column gives the fraction of the volume within each survey's observed footprint that is excluded, and the second gives the fraction of the full 90\% localisation volume excluded, accounting for each survey's sky coverage. For each pointing direction, only the most constraining observation is used (Section~\ref{sec:coverage_calc}).}
    \label{tab:volume_excluded}
\end{table*}

\subsubsection{ATLAS}
\label{sec_sub:ATLAS_Coverage}
ATLAS observed 87\% of the skymap to a median depth of 19.3, 19.4, and 19.7\,mag in the $o$, $c$, and $w$ bands, respectively. ATLAS observations in this analysis span 0.2 to 14 days post-merger. 

\begin{figure*}
    \centering
    \begin{subfigure}{0.95\textwidth}
        \centering
        \includegraphics[width=\linewidth]{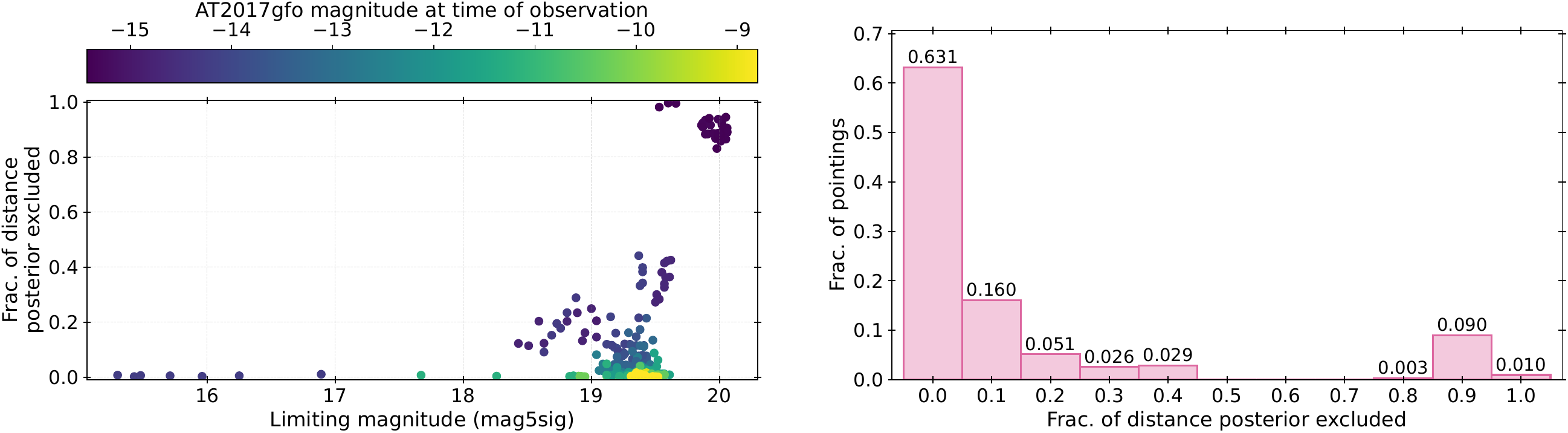}
        \vspace{0.3cm}
    \end{subfigure}
    \begin{subfigure}{0.95\textwidth}
        \centering
        \includegraphics[width=\linewidth]{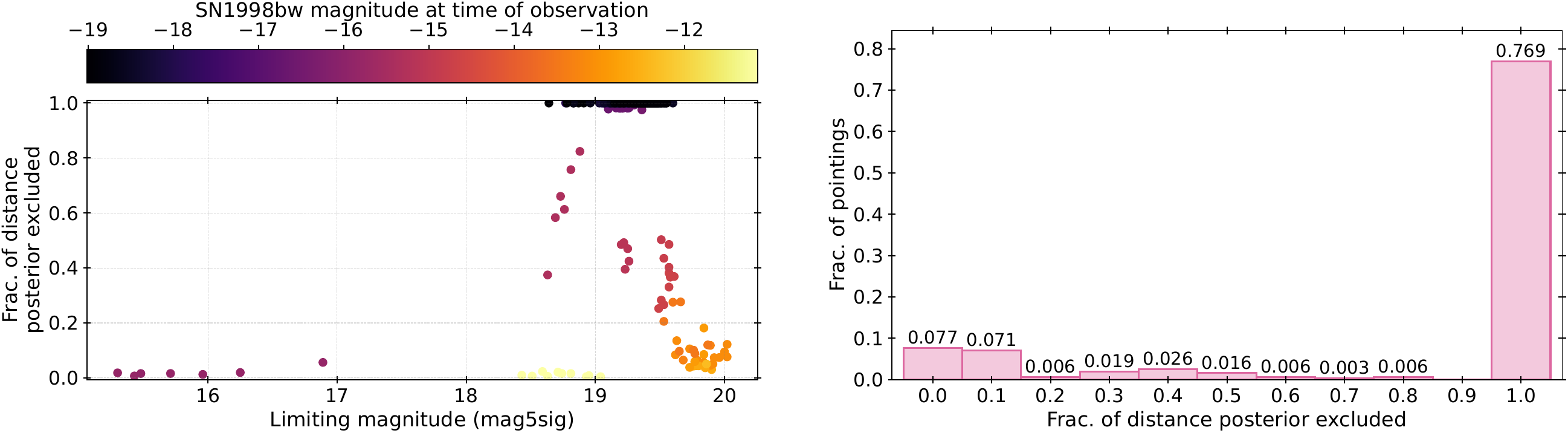}
    \end{subfigure}
    
    \caption{Volume exclusion analysis for the ATLAS observations of S251112cm, (top) the AT~2017gfo-like kilonova and (bottom) the SN~1998bw-like Type Ic-BL model. Left: each point is a single pointing, plotted by its $5\sigma$ limiting magnitude against the fraction of the LVK distance posterior along that line of sight within which the model would have been detected. Points are coloured by the model absolute magnitude at the epoch of observation. A value of 1 indicates the model would be detected out to the maximum LVK distance. Right: distribution of the excluded fraction across all pointings, retaining only the most constraining observation for each pointing direction.}
    \label{fig:exclusion_analysis_ATLAS}
\end{figure*}

The fraction of probability ruled out for each pointing, a calculation that we term ``exclusion analysis'', for these observations can be seen in Fig. \ref{fig:exclusion_analysis_ATLAS}. Figure \ref{fig:exclusion_analysis_ATLAS}(top) shows that for the majority of ATLAS pointings we can exclude little of the distance range for an AT~2017gfo-like kilonova. A small group of the deepest pointings, reaching $\sim20$\,mag, is very constraining for certain sight lines. These observations were taken between 0.2 and 0.7 days post-merger, when the kilonova model is still near peak at $\sim -15$\,mag. This behaviour is summarised in the right-hand panel which shows that 63\% of the ATLAS pointings would not have seen a kilonova and only 10\% would have seen it in more than 90\% of the distance posterior.

Integrated over the entire skymap, the ATLAS observations rule out an AT~2017gfo-like kilonova in 14\% of the \emph{observed} volume, or 12\% of the entire 90\% localisation volume (Table \ref{tab:volume_excluded}). 

For the scenario of a SN~1998bw-like counterpart, shown in Fig. \ref{fig:exclusion_analysis_ATLAS}(bottom), the same images are far more constraining. The observations cluster into different nights, depending on the brightness predicted by the model; those where the model was fainter than approximately $-14$\, mag exclude little of the distance posterior, while a much greater fraction (77\%) of pointings would have seen a SN~1998bw-like SN Ic-BL in 100\% of the distance posterior. This rules out a SN~1998bw-like SN Ic-BL over $\sim80$\% of the observed ATLAS volume and 70\% of the entire 90\% localisation region. 

\begin{figure*}
    \centering
    \begin{subfigure}{0.95\textwidth}
        \centering
        \includegraphics[width=\linewidth]{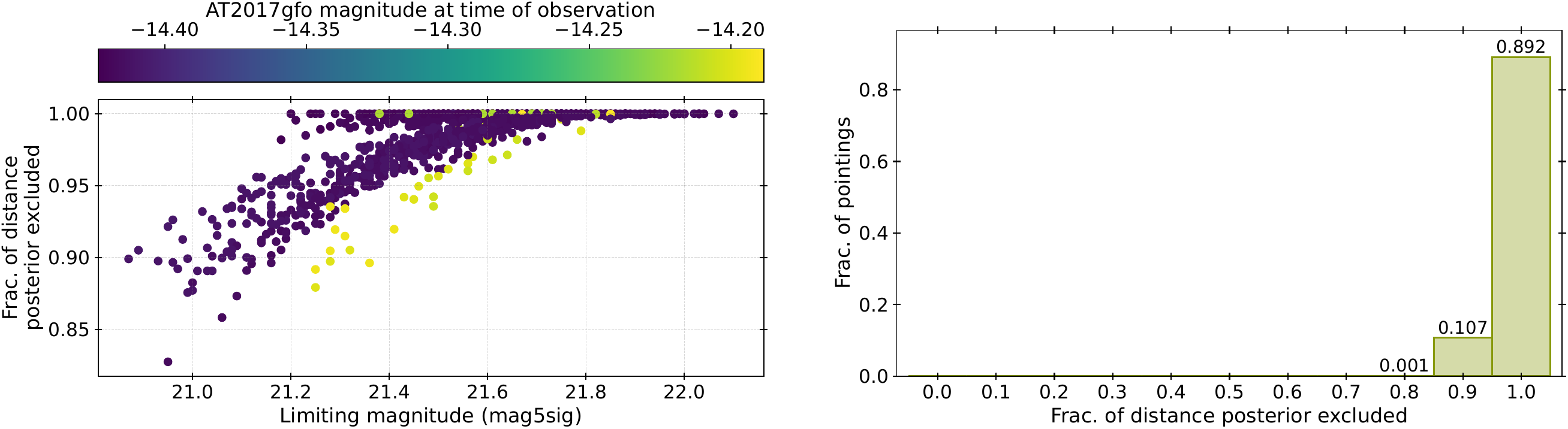}
        \vspace{0.3cm}
    \end{subfigure}
    \hfill
    \begin{subfigure}{0.95\textwidth}
        \centering
        \includegraphics[width=\linewidth]{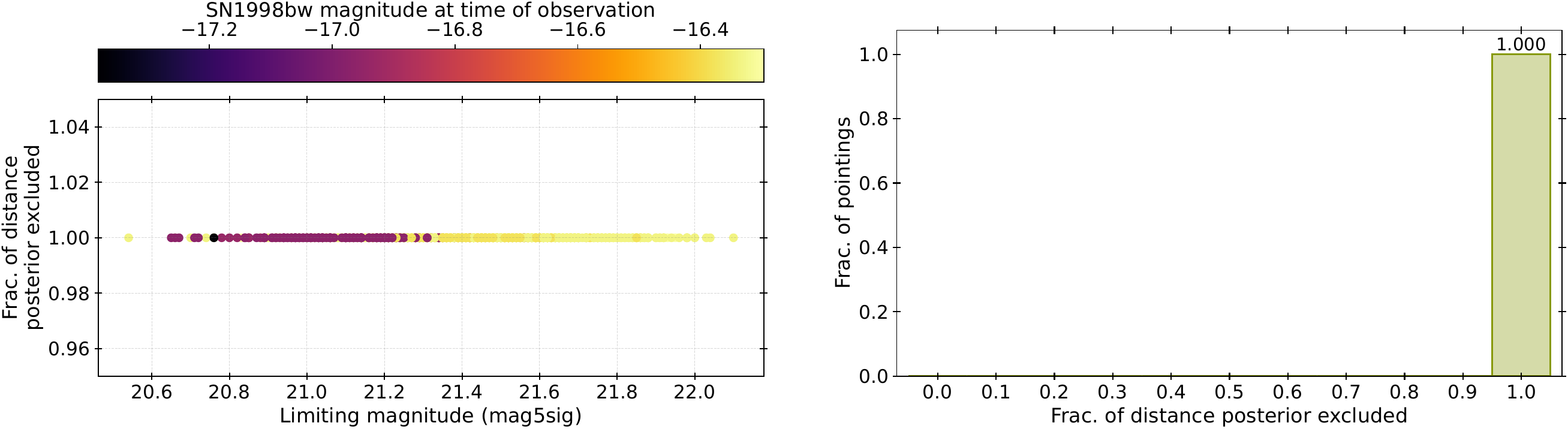}
    \end{subfigure}
    \caption{As Fig. \ref{fig:exclusion_analysis_ATLAS}, but for the Pan-STARRS observations. For the SN~1998bw-like model, every pointing excludes the full LVK distance posterior.}
    \label{fig:exclusion_analysis_Pan-STARRS}
\end{figure*}

\begin{figure*}
    \centering
    \begin{subfigure}{0.95\textwidth}
        \centering
        \includegraphics[width=\linewidth]{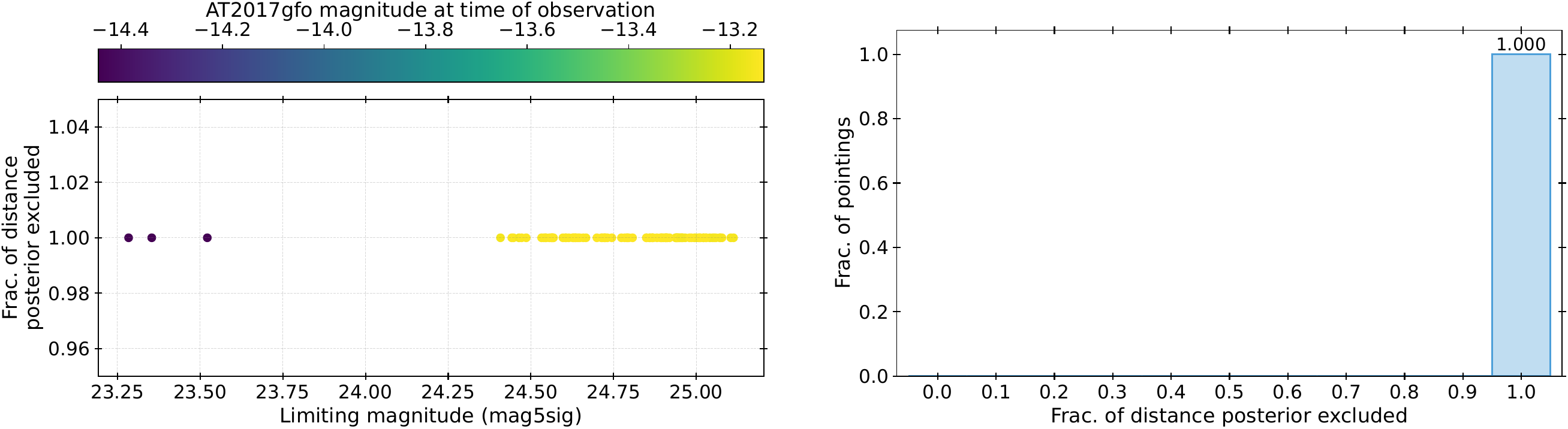}
        \vspace{0.3cm}
    \end{subfigure}
    \hfill
    \begin{subfigure}{0.95\textwidth}
        \centering
        \includegraphics[width=\linewidth]{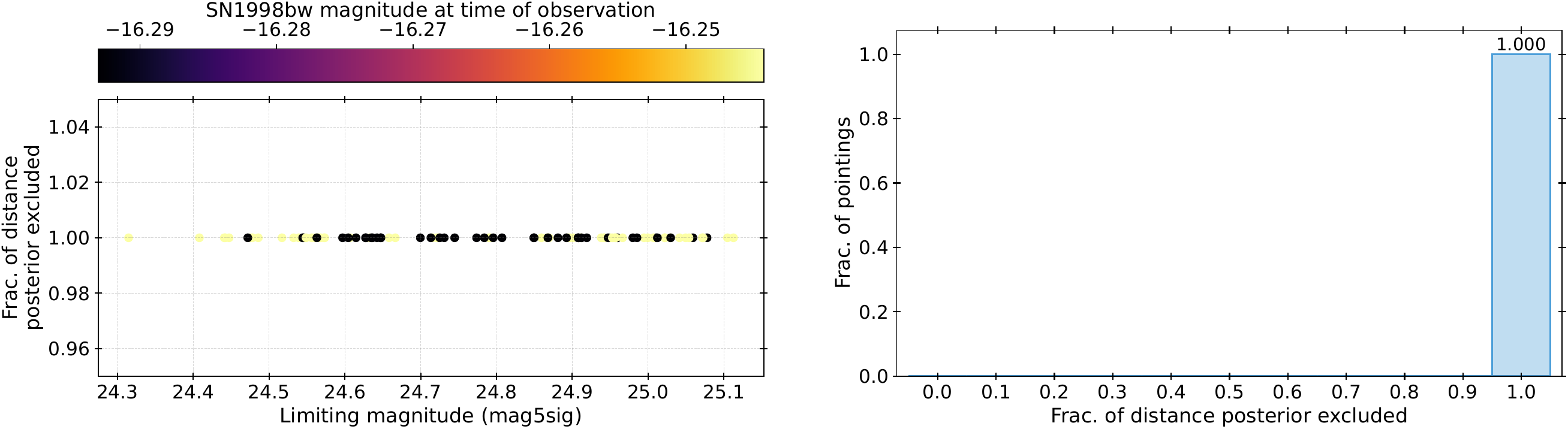}
    \end{subfigure}
    \caption{As Fig. \ref{fig:exclusion_analysis_ATLAS}, but for the Rubin ToO observations. For both models, every pointing excludes the full LVK distance posterior.}
    \label{fig:exclusion_analysis_Rubin}
\end{figure*}

\subsubsection{Pan-STARRS}
\label{sec_sub:Pan-STARRS_Coverage}
Pan-STARRS covered 64.7\% of the 90\% localisation region reaching median depths of 21.1 and 21.9\,mag in the \ips, and \wps\ bands between 2.6--7 days post-merger.

The excluded localisation volume analysis for Pan-STARRS mirrors the method used for ATLAS and is shown in Fig. \ref{fig:exclusion_analysis_Pan-STARRS}. The greater depth of Pan-STARRS is immediately apparent, with the limiting magnitudes ranging from $\sim21 -22$ mag. For the AT~2017gfo-like kilonova model shown in Fig. \ref{fig:exclusion_analysis_Pan-STARRS} (top), for all pointings we are able to exclude a kilonova over at least 82\% of the distance posterior, and for observations reaching limiting magnitudes of $\sim22$ we can completely rule out a kilonova at that position. 89\% of the Pan-STARRS observations exclude the full LVK distance posterior for the AT~2017gfo-like kilonova model. Across all Pan-STARRS pointings, the observations ruled out an AT~2017gfo-like kilonova across 99\% of the observed volume, and across 64\% of the entire 90\% LVK localisation volume (Table \ref{tab:volume_excluded}).    

For a SN~1998bw-like transient, shown in Fig. \ref{fig:exclusion_analysis_Pan-STARRS}(bottom), we can exclude the complete distance posterior at every observed pointing. As a result the histogram shows a single bar at 1 which represents 100\% of the Pan-STARRS volume and 65\% of the entire localisation region ruling out a SN~1998bw-like Type Ic-BL. Unlike ATLAS, the limiting factor for the Pan-STARRS observations is no longer the depth, but the fraction of the skymap that could be observed from the northern locations of the telescopes. 

\subsubsection{Vera C. Rubin}
\label{sec_sub:Rubin_Coverage}
Rubin observed 35.73\% and 35.90\% of the skymap to a median depth of 24.46 and 23.53\,mag in the $g$ and $i$  bands, respectively. These observations span 2 to 10 days post-merger and assume a 1.75$^\circ$ vignetting radius (the radius used for prompt processing), and no detector 122 (which has been powered off since April 2024).

The volume excluded by Rubin ToO observations can be seen in Fig. \ref{fig:exclusion_analysis_Rubin}. In contrast to both ATLAS and Pan-STARRS, the Rubin pointings are never depth-limited for either model. For the AT~2017gfo-like kilonova model, Fig. \ref{fig:exclusion_analysis_Rubin}(top), the most constraining observation of each pointing spans limiting magnitudes of 23.3--25.0 mag, with each pointing excluding the full LVK distance posterior. The histogram therefore reduces to a single bar at 1. This same behaviour can be seen for the SN~1998bw-like Type Ic-BL model, Fig. \ref{fig:exclusion_analysis_Rubin}(bottom). 

Across all Rubin pointings, the observations ruled out both an AT~2017gfo-like kilonova and SN~1998bw-like Ic-BL across 100\% of the observed volume, corresponding to 36\% of the entire LVK 90\% volume (Table \ref{tab:volume_excluded}) -- essentially, the fraction of the 90\% area that Rubin observed \citep{MacBride2026}. 

As with Pan-STARRS, the most limiting factor for Rubin is not depth, but the fraction of skymap covered. This follows from the decision to target the 70\% region (Section \ref{sec_sub:rubin_obs}), and from the southern accessibility of Cerro Pach\'on. The absence of any depth dependency indicates the capability of detecting much fainter counterparts, quantified in Section \ref{sec_sub:coverage_calc_vary_mag}.

\subsubsection{Collective survey coverage}
\label{sec_sub:Collective_Coverage}
\begin{figure*}
    \centering

    \includegraphics[width=0.8\linewidth]{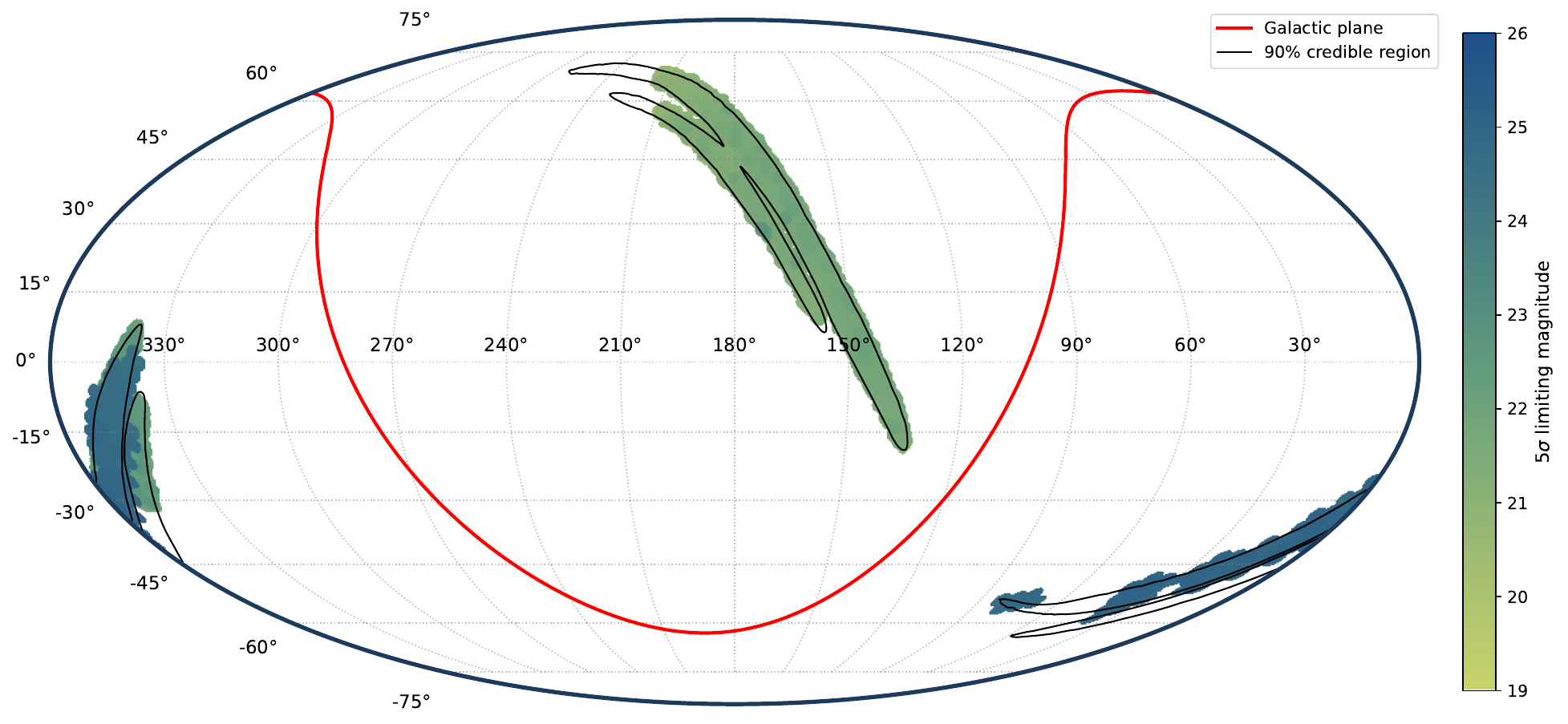}
    \caption{Combined Pan-STARRS and Rubin coverage of the S251112cm 90\% credible region (black contour), coloured by the deepest $5\sigma$ limiting magnitude achieved at each position. Together the two surveys cover 1387.5\,deg$^2$ ($\sim80\%$) of the 1681.1\,deg$^2$ 90\% region; Rubin dominates in the south and Pan-STARRS in the north. The red line traces the Galactic plane.}
    \label{fig:Rubin_PS_Depth}
\end{figure*}

Figure \ref{fig:Rubin_PS_Depth} shows the combined Pan-STARRS and Rubin coverage of the \texttt{bilby.multiorder.fits} 90\% credible region. Together the two surveys cover a total of 1387\,deg$^2$ of the 1681\,deg$^2$, or approximately 80\%, of the 90\% credible region leaving only 294\,deg$^2$ ($\sim$20\%) unobserved by either facility. 

Combining the volume excluded by Pan-STARRS and Rubin in Sections \ref{sec_sub:Pan-STARRS_Coverage} and \ref{sec_sub:Rubin_Coverage}, we jointly rule out an AT~2017gfo-like kilonova and a SN~1998bw-like Type Ic-BL across $\sim80\%$. Since both telescopes achieve the necessary depth to detect a kilonova over the majority of their pointings, the residual volume is set almost entirely by the unobserved $\sim20\%$ of the sky area, rather than sensitivity.

For comparison, ATLAS covered 1545\,deg$^2$ of the 90\% credible region, a wider area than Pan-STARRS and Rubin combined, but to a shallower depth ($m \sim 19.5$). For an AT~2017gfo-like kilonova at the distance of S251112cm this depth is the limiting factor: the model falls below the ATLAS limit across most of the distance posterior, and ATLAS excludes it within only 14\% of its observed volume, or 12\% of the full 90\% localisation volume. A SN~1998bw-like supernova is substantially brighter, and for a source that lies above the limiting magnitude across most of the distance posterior, sky coverage rather than depth determines the constraint. ATLAS rules out a SN~1998bw-like counterpart across 70\% of the 90\% localisation volume, more than either Pan-STARRS (65\%) or Rubin (36\%) individually.

\subsubsection{Varying magnitudes of models}
\label{sec_sub:coverage_calc_vary_mag}
\begin{figure*}
    \centering
    \begin{subfigure}{0.95\textwidth}
        \centering
        \includegraphics[width=\linewidth]{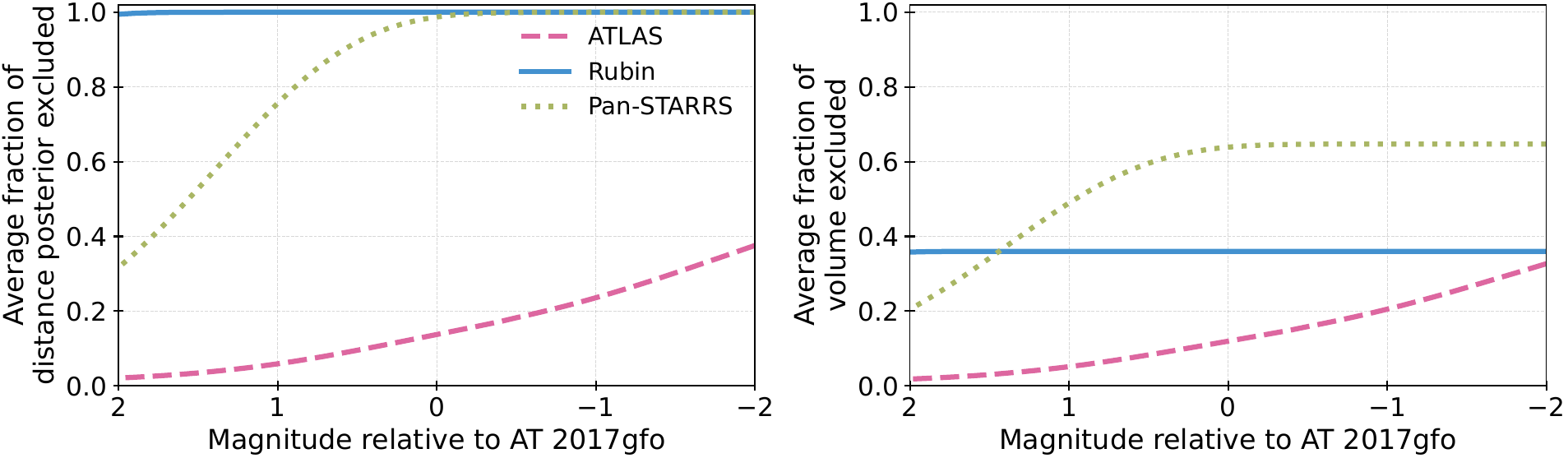}
        \vspace{0.5cm}
    \end{subfigure}
    \hfill
    \begin{subfigure}{0.95\textwidth}
        \centering
        \includegraphics[width=\linewidth]{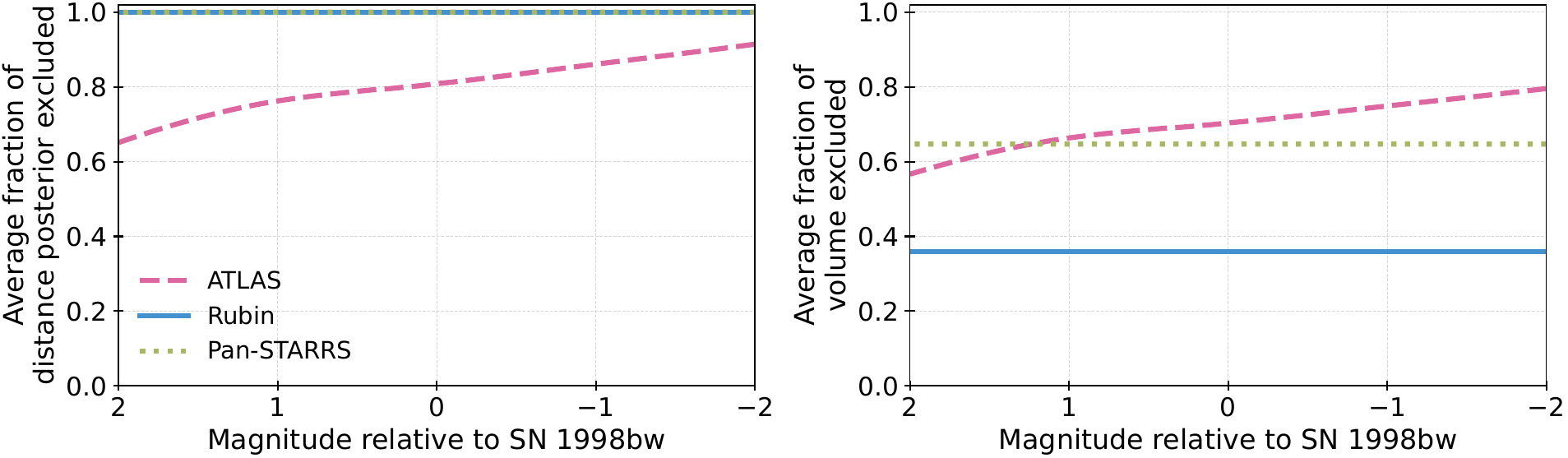}
    \end{subfigure}
    \caption{Sensitivity of the volume exclusion constraints to model brightness, for the AT~2017gfo-like kilonova (top) and the SN~1998bw-like Type Ic-BL model (bottom), with uniform magnitude offsets of $-2$ to $+2$ mag applied to the model light curves (positive values are fainter). Left: mean fraction of the LVK distance posterior excluded per pointing within each survey's footprint, a measure of depth. Right: fraction of the full 90\% localisation volume excluded by each survey, which additionally accounts for sky coverage. ATLAS is denoted using the dashed magenta, Rubin the solid blue, and Pan-STARRS the dotted green line.}
    \label{fig:varying_mag_plots}
\end{figure*}

All constraints mentioned above assume the model light curves of Section \ref{secsub:MOSFiT_Models} at their nominal brightness. In practice the luminosity of any counterpart is highly uncertain. Kilonovae associated with short GRBs span a range of luminosities around AT~2017gfo \citep{Gompertz2018_KN_Diversity}, and the ejecta mass and composition of a merger involving a subsolar-mass compact object are essentially unconstrained, so a counterpart to S251112cm could be considerably brighter or fainter than our models. To test how they depend on this assumption, we repeated the volume exclusion analysis after applying uniform offsets of $\pm2$ mag to each model light curve. Figure \ref{fig:varying_mag_plots} shows the results. The left-hand panels give the fraction of the distance posterior excluded within each survey's obtained footprint, which measures sensitivity. The right-hand panels give the fraction of the full 90\% volume excluded by each survey. 

For the AT~2017gfo-like kilonova, \ref{fig:varying_mag_plots}(top), the Rubin constraint remains the same across the explored magnitude range, excluding 99\% of the distance posterior within its footprint even at $+2$ mag. Therefore, Rubin's volume constraint also remains at 36\% across the magnitude range. Pan-STARRS excludes essentially the full posterior within its footprint for kilonova counterparts up to $\sim0.5$ mag fainter than AT~2017gfo, beyond which the excluded fraction falls to $\sim 75\%$ at $+1$ mag but remains as high as $\sim 30\%$ even at $+2$ mag. As a result, the volume constraint drops from 64\% to $\sim20\%$. 

A uniform offset changes the brightness of the model but not the shape of its light curve. A kilonova with lower ejecta mass would be fainter, but would also have a shorter photon diffusion time, causing it to peak earlier and fade more rapidly \citep{Metzger2017_kilonovae}. A fixed magnitude offset from AT~2017gfo is therefore not a direct proxy for lower ejecta mass mergers. Even so, at the distance of S251112cm, the depth of Rubin observations is unlikely to be the limiting factor, as \citet{Andreoni2024} show that at nominal depths, the red emission from a kilonova with an order of magnitude less ejecta than AT~2017gfo is within reach of a single Rubin visit at 200\,Mpc, roughly twice the distance of S251112cm.  

For the SN~1998bw-like Type Ic-BL, shown in Fig. \ref{fig:varying_mag_plots}(bottom), both Pan-STARRS and Rubin exclude the full distance posterior within their footprints across the entire $\pm2$ mag range. Therefore, the Pan-STARRS and Rubin constraints on the 90\% localisation volume remain constant at 65\% and 36\%. ATLAS is again the only survey whose constraint depends on brightness over this range, falling from $\sim80\%$ to $\sim 55\%$ of the localisation volume between $\pm2$ mag. However, with its wide coverage over both hemispheres, it excludes a comparable volume to Pan-STARRS for a Type Ic-BL that is 1\,mag fainter than SN~1998bw.

\section{Discussion}
\label{Disussion}

After extensive vetting of candidates, we concluded that neither of the two EM counterparts proposed in this work for SSM mergers -- a kilonova or a Type Ic-BL supernova -- was detected during this campaign. We briefly summarise here our observational constraints and those from other teams, before discussing the possible implications for progenitor channels.

Section \ref{sec_sub:coverage_calc_vary_mag} quantifies how far our AT~2017gfo constraint extends to fainter counterparts. Over the Rubin footprint the excluded fraction for a kilonova two magnitudes fainter than AT~2017gfo is 99\%, whereas the Pan-STARRS constraint begins to decline within half a magnitude, dropping to 31\% at two magnitudes fainter. This means that while Pan-STARRS and Rubin rule out approximately 80\% of the 90\% localisation volume for an AT~2017gfo-like counterpart, only 55\% of the localisation volume is excluded at two magnitudes fainter. Whether a kilonova could have escaped detection therefore depends on its luminosity, which is set largely by the mass of radioactive material ejected.

The ejecta mass of a NS merger involving one or two SSM neutron stars has not been measured, and such mergers have only recently been simulated. A binary of lower total mass might naively be expected to eject less material than AT~2017gfo, and so produce a fainter transient. 

The fitting formulae used to map binary parameters to ejecta masses, such as the dynamical ejecta relation of \citet{Kruger2020} adopted in the Nuclear-physics and Multi-Messenger Astrophysics framework NMMA \citep{Pang2023}, are calibrated on simulations of binaries with components above $\sim1\,\text{M}_\odot$, and their behaviour at low-compactness is an extrapolation. In the \citet{Kruger2020} relation, the term inversely proportional to compactness has a negative coefficient and dominates for low-compact stars. Hence the fit predicts little to no dynamical ejecta for near-equal-mass SSM binaries, the opposite of the trend found in the simulations. 

However, the available models point the other way. In the first simulations of SSM binary neutron star mergers, the dynamical ejecta mass increases as the stars become lighter and more deformable.  In the merger simulations of \citet{Gao2026}, the dynamical ejecta mass increases from $5.5\times10^{-3}$ to $1.0\times10^{-2}\,\text{M}_{\odot}$ as the mass of each neutron star in an equal-mass binary decreases from 0.7 to $0.4\,\text{M}_{\odot}$. Unequal-mass binaries eject more still, both \citet{Gao2026}, for mass ratios of 1.22--1.5, and \citet{Corman2026}, for a subsolar-mass star paired with a heavier neutron star, find dynamical ejecta masses of a few $\times10^{-2}\,\text{M}_{\odot}$.

The disc outflows, which dominate the ejecta budget in most models, may also be substantial. The disc-mass relation used in NMMA \citep{Dietrich2020, Pang2023} depends on the ratio of the total binary mass to the threshold mass for prompt collapse \citep{Agathos2020}, which for any plausible EOS lies well above the total mass of two SSM neutron stars. Such a binary merger remnant does not collapse promptly; a longer-lived remnant will irradiate the disc outflow with neutrinos, raising its electron fraction and making the emission brighter and bluer \citep{Metzger2014}. These relations are only indicative at the lowest masses allowed for S251112cm, since no simulations yet extend below $0.4\,\text{M}_\odot$ per component \citep{Gao2026}. However, when taken together the simulations and disc-mass relation do not indicate that the total ejecta mass should fall well below that inferred for GW170817 ($\sim 0.006\,\text{M}_\odot$ dynamical and $\sim 0.07\,\text{M}_\odot$ disc wind; \citealt{Pang2023}). 

Other campaigns probing S251112cm include \citet{Vieira2026_TROVE} who vetted all 456 candidates reported by the community, identifying no likely counterpart, \citet{Hall2026_ZTF_S251112cm} reporting ZTF, FTW, and DECam coverage of the localisation region, and \citet{Paek2026} monitoring the highest-ranked host galaxies in the NIR for two months which probed the longer lived emission predicted in some SSM formation channels. 

While S251112cm has a near 100\% probability that at least one merger component is subsolar, the available information regarding chirp mass simply states that it lies in the range 0.10--0.87\,$\text{M}_\odot$ \citep{LVGCNdiscovery}. At the upper end of this range, the primary may be sub- or super-solar. A merger of two SSM objects with equal masses (mass ratio, $q$ = 1) is compatible. However, a primary of $\sim 1 - 3.5\,\text{M}_\odot$ is possible for $q=0.1$ \citep[the limit of the available SSM template bank;][]{Hanna2025}; see also \citet{Vieira2026_TROVE}. Although, we note that LVK reported a probability of only 0.08 that the system contains a 1--3\,$\text{M}_\odot$ object. 

Two classes of binary could produce the S251112cm signal while containing a neutron star. Either a SSM neutron star merging with a heavier black hole, or a SSM black hole merging with a neutron star of standard mass. Here we consider both. 

In the first case, the low compactness expected of a SSM neutron star places the binary inside of the tidal disruption regime. For a $0.35\,\text{M}_\odot$ neutron star and a non-spinning $3.5\,\text{M}_\odot$ black hole, the fitting formula of \citet{Foucart2018} predicts that mass remains outside the black hole after merger. This formula, calibrated against numerical-relativity simulations, gives the total baryonic mass remaining outside the black hole as a function of the symmetric mass ratio, the black hole spin and the neutron star compactness. The remnant mass increases as compactness decreases, so the low compactness of a SSM neutron star favours tidal disruption.

In the second case, \citet{Markin2023} simulated the merger of a $0.5\,\text{M}_\odot$ black hole with a $1.4\,\text{M}_\odot$ neutron star, finding $\sim0.04\,\text{M}_\odot$ of lanthanide-rich dynamical ejecta. The resulting kilonova peaks in the infrared at $M\sim-15$\,mag after 2--3 days, but is considerably fainter in the optical, at $M\sim-12$ to $-13$\,mag one day after merger, with up to a 2\,mag difference depending on viewing angle \citep{Markin2023}. Their models include only the dynamical ejecta, so additional, potentially bluer, post-merger components are not captured. In the optical, a transient would be approximately 3 -- 4\,mag fainter than AT~2017gfo. At the median LVK distance of 93\,Mpc, it would have an apparent magnitude of approximately 22--23\,mag one day after merger, at or below the median Pan-STARRS $w$ depth and well beyond that of ATLAS. The Rubin depths of $m_g \simeq 24.5$ and $m_i \simeq 23.5$ are in principle sensitive to this emission. However, Rubin observations began two days after merger, later than the epoch at which \citet{Markin2023} quote the optical magnitudes, and covered only 36\% of the localisation region. Our data therefore place only weak constraints on this scenario.

In the collapsar channel, SSM neutron stars form in the gravitationally unstable accretion disc surrounding the black hole, and undergo mergers within hours to a number of days \citep{Metzger2024_SSM_Ic_BL, Chen_and_Metzger_SSM_Ic_BL}. The merger would be accompanied by a supernova that is longer lived and more luminous than a kilonova. Every Rubin and Pan-STARRS pointing rules out a SN~1998bw-like Type Ic-BL supernova across the full LVK distance posterior for every observation. Jointly, Pan-STARRS and Rubin exclude a SN~1998bw-like counterpart across $\sim80$\% of the LVK volume. Should S251112cm have originated from a collapsar disc, the accompanying supernova would have been recovered unless it was located in the remaining $\sim20$\% of unobserved area.  This channel also predicts further gravitational wave signals, as merger remnants in the disc inspiral into the central black hole \citep{Metzger2024_SSM_Ic_BL}. No additional GW signals were reported alongside S251112cm (or S250818k). However, it is also worth noting that there was a period of detector downtime that lasted 42 hours, which started $\sim 15$ hours after S251112cm, in which these additional signals might have gone undetected \citep{Vieira2026_TROVE, Hall2026_ZTF_S251112cm}. 

Our non-detections in the optical could also point to a channel with no electromagnetic counterpart. While standard stellar evolution cannot produce SSM black holes, primordial black holes remain a possibility \citep{Carr2021}. A merger of two primordial black holes is not expected to produce any counterpart outside gas-rich environments. Even in a gas-rich environment, the S251112cm component masses are $\sim10^2$ times lower than those of GW190521, the merger which has been proposed to be associated with a candidate AGN flare \citep{Graham2020}. \citet{Vieira2026_TROVE} highlight that this mass difference could result in a flare $\sim10^4$ times less luminous. \citet{Haque2026} examine the primordial black hole interpretation of S251112cm directly, and find that if primordial black holes in the $\sim0.5$--$1\,\text{M}_{\odot}$ range exist, there is a reasonable probability that LVK would have detected one at its current sensitivity. A primordial origin therefore remains a viable explanation for both the signal and the absence of a counterpart if the GW signal is confirmed as real.

Taken together, the wide field data leave little room for an AT~2017gfo-like kilonova or a SN~1998bw-like supernova from S251112cm. If there was such an EM counterpart to S251112cm, it would have to have occurred in the unobserved $\sim$20\% of the localisation volume. Even then, neither the isolated low-mass merger model (for stellar evolution reasons) nor collapsar disc fragmentation model (due to the lack of further GW signals) is fully satisfactory in explaining S251112cm. Alternative possibilities remain speculative: a merger involving at least one SSM (possibly primordial) black hole \citep{Markin2023}, or an accretion disc in the Type IIb SN~2025adtq \citep{Hall2026_ZTF_S251112cm}.

\section{Conclusions}
\label{sec:conclusions}
In this paper we have presented and analysed wide-field optical follow-up of S251112cm, a SSM merger candidate detected by the LVK network at $93 \pm 27$\,Mpc with a false alarm rate of one per four years. Pan-STARRS covered 1084\,deg$^2$ (65\%) of the 90\% localisation region within 7 days, ATLAS 1545\,deg$^2$ (87\%) within 14 days, and the Vera C. Rubin Observatory 849\,deg$^2$ (36\%) between 2 and 10 days post-merger. Pan-STARRS and Rubin together reached 1387\,deg$^2$ of the 1681\,deg$^2$ 90\% credible region ($\sim80$\%), leaving 294\,deg$^2$ unobserved by either facility. To our knowledge this is the deepest and most complete optical coverage of a SSM candidate, and among the most complete for any potentially EM-bright GW source to date.

The three surveys discussed in this manuscript reported 269 unique candidates collectively. Vetting on prior activity, host luminosity distance, photometric evolution, and absolute magnitude left eight, seven of which had distances that were unconstrained. Follow-up VLT/FORS2 spectroscopy places AT~2025alqe beyond the GW localisation volume. All candidate hosts with Pan-STARRS1 photometric redshifts \citep{Tarrio2020} lie beyond the $3\sigma$ LVK distance posterior. The remaining candidate, AT~2025aimy, has a host at a spectroscopic distance of 140\,Mpc consistent with the GW volume, but its light curve evolves too slowly for our kilonova model, fading by only 0.83\,mag in the $g$ band between 3 and 10\,d compared with 4.38\,mag for AT~2017gfo. Furthermore, its host luminosity lies in the faintest 1\% of the ATLAS100 sample of transient host galaxies, an unlikely environment for a rare kilonova. As a result, we find no convincing counterpart.

At median $5\sigma$ depths of 19.3\,mag (ATLAS $o$), 21.9\,mag (Pan-STARRS $w$),
and 24.5 and 23.5\,mag (Rubin $g$, $i$), Pan-STARRS and Rubin together exclude both an AT~2017gfo-like kilonova and SN~1998bw-like Type Ic-BL supernova over approximately 80\% of the 90\% localisation region. The AT~2017gfo and SN~1998bw models are excluded in 99\% and 100\% of the distance posterior within the combined Pan-STARRS and Rubin footprints. 

It is therefore unlikely, though not impossible, that S251112cm produced a counterpart as bright as an AT\,2017gfo-like kilonova or a collapsar-driven Type Ic-BL supernova. S251112cm demonstrates the capability of the current wide-field survey network. In combination with existing facilities, Rubin can now provide near-complete hemispheric coverage, surveying the southern sky (to deep limits, $r \sim$ 24\,mag) while Pan-STARRS covers the north (down to Dec\,$=-30^\circ$) with typical single-visit depths of $\rps \sim 21.8$\,mag. Collectively, wide-field imaging and high-cadence surveying represent a significant step forward in the efficiency and completeness of EM follow-up of GW events, and this suite will play a fundamental role in the LVK Collaboration's forthcoming observing runs. O4 (May 2023 -- November 2025) already enabled BNS detections out to 150--160\,Mpc \citep{Arnaud2026_LVK_O4}. The Intermediate Observing Run (IR1), planned for early 2027, is expected to reach 160--180\,Mpc, and will be followed by O5, which has no confirmed start date but is expected to reach substantially greater sensitivity\footnote{As of 2026 Nov, see: \href{https://observing.docs.ligo.org/plan/}{https://observing.docs.ligo.org/plan/}}. The corresponding increase in detection distance means that EM counterparts will be systematically fainter and more difficult to identify, further emphasising the need for deep, wide-field facilities such as Rubin to efficiently probe these more distant mergers.

\section*{Acknowledgements}
 
AJS thanks Lynne Jones for guidance on retrieving the Rubin ToO observations of S251112cm from the pre-LSST visit database and Early Data Preview 2. AJS also thanks Aleksandra Bochenek, Ernst de Mooij and Christopher Watson for helpful discussions, and acknowledges travel support provided by STFC for UK participation in LSST through grant ST/X001334/1. 
MN is supported by the Science and Technology Facilities Council (grant number UKRI4107) and by the European Research Council (grant agreement No.~948381).
T.-W.C. acknowledges support from the Ministry of Education, Taiwan, through the Yushan Fellow Program (MOE-111-YSFMS-0008-001-P1), and the National Science and Technology Council (NSTC 114-2112-M-008-021-MY3).
KWS acknowledges funding from the Royal Society.

Pan-STARRS is primarily funded to search for near-earth asteroids through NASA grants NNX08AR22G and NNX14AM74G. The Pan-STARRS science products for LIGO-Virgo-KAGRA follow-up are made possible through the contributions of the University of Hawaii's Institute for Astronomy and Queen's University Belfast.

The Pan-STARRS1 Sky Surveys have been made possible through contributions by the University of Hawaii's Institute for Astronomy, the Pan-STARRS Project Office, the Max Planck Society and its participating institutes, the Max Planck Institute for Astronomy, Heidelberg and the Max Planck Institute for Extraterrestrial Physics, Garching, Johns Hopkins University, Durham University, the University of Edinburgh, Queen's University Belfast, the Harvard-Smithsonian Center for Astrophysics, the Las Cumbres Observatory Global Telescope Network Incorporated, the National Central University of Taiwan, the Space Telescope Science Institute, and the National Aeronautics and Space Administration under Grant No.~NNX08AR22G issued through the Planetary Science Division of the NASA Science Mission Directorate, the National Science Foundation Grant No.~AST-1238877, the University of Maryland, Eotvos Lorand University, and the Los Alamos National Laboratory.

This work has made use of data from the Asteroid Terrestrial-impact Last Alert System project. ATLAS is primarily funded to search for near-earth asteroids (NEOs) through NASA grants NN12AR55G, 80NSSC18K0284 and 80NSSC18K1575; by-products of the NEO search include images and catalogs from the survey area.  The ATLAS science products have been made possible through the contributions of the University of Hawaii's Institute for Astronomy, Queen's University Belfast, the Space Telescope Science Institute and the South African Astronomical Observatory.

This material is based upon work supported in part by the National Science Foundation through Cooperative Agreements AST-1258333 and AST-2241526 and Cooperative Support Agreements AST-1202910 and 2211468 managed by the Association of Universities for Research in Astronomy (AURA), and the Department of Energy under Contract No. DE-AC02-76SF00515 with the SLAC National Accelerator Laboratory managed by Stanford University. Additional Rubin Observatory funding comes from private donations, grants to universities, and in-kind support from LSST-DA Institutional Members.

Based on observations collected at the European Organisation for Astronomical Research in the Southern Hemisphere under ESO programme 117.2AX3. 

The Photometric Redshifts for the Legacy Surveys (PRLS) catalog used in this paper was produced thanks to funding from the U.S. Department of Energy Office of Science, Office of High Energy Physics via grant DE-SC0007914.

\section*{Data Availability}
The LVK skymap (\texttt{Bilby.multiorder.fits,1}) for S251112cm is publicly available from GraceDB (\url{https://gracedb.ligo.org/superevents/S251112cm}). All candidates discussed in this work were reported to the Transient Name Server (\url{https://www.wis-tns.org}) and in the GCN Circulars cited in the text. ATLAS photometry is available through the ATLAS Forced Photometry Server (\url{https://fallingstar-data.com/forcedphot/}). Host-galaxy images, photometry and photometric redshifts were taken from the DESI Legacy Imaging Surveys (\url{https://www.legacysurvey.org}; \citealt{Zhou2021,Zhou2023}), the Pan-STARRS1 DR2 catalogue available through MAST (\url{https://catalogs.mast.stsci.edu/panstarrs/}), and the Pan-STARRS1 photometric redshift catalogue of \citet{Tarrio2020}. Rubin Observatory visit data were obtained from the pre-LSST visit database and Early Data Preview 2, and are available to individuals with Rubin data rights. The VLT/FORS2 spectrum (ESO programme 117.2AX3) will be available from the ESO Science Archive after the proprietary period.



\bibliographystyle{mnras}
\bibliography{example} 




\appendix


\bsp	
\label{lastpage}
\end{document}